\documentclass{aa}
\usepackage{graphicx}
\usepackage[normalem]{ulem}
\newcommand{\cmtwo}{\,{\rm cm^{-2}}}

\newcommand{\ergs}{\,{\rm erg\, s^{-1}}}
\newcommand{\pc}{\,{\rm pc}}
\newcommand{\kpc}{\,{\rm kpc}}
\newcommand{\m}{\,{$\mu$m}}

\begin{document}
   \title{Dissecting the spiral galaxy M\,83: mid-infrared emission
 and comparison with other tracers of star formation\thanks{Based on observations with ISO, an ESA project with instruments
funded by ESA Member States (especially the PI countries: France, Germany, the
Netherlands and the United Kingdom) and with the participation of ISAS and NASA.} }

   \author{A.~Vogler\inst{1}, S.~C.~Madden\inst{1}, R.~Beck\inst{2},
	    A.A. Lundgren\inst{3,4}, M.~Sauvage\inst{1}, L.~Vigroux\inst{1}
           \and M.~Ehle\inst{5}}
   \institute{DSM/DAPNIA/Service d'Astrophysique, CEA/Saclay, 91191,
   Gif-sur-Yvette Cedex, France
   \and
   Max-Planck-Institut f\"ur Radioastronomie, Auf dem H\"ugel 69, 53121 Bonn, Germany
   \and
European Southern Observatory, Casilla 19001, Santiago 19, Chile
   \and
   Stockholm Observatory, AlbaNova, 106 91 Stockholm, Sweden
   \and
   XMM-Newton Science Operations Centre,
   European Space Agency, Villafranca, PO Box 50727, 28080 Madrid, Spain
}

\date{Accepted July 7, 2005}

\offprints{S.C.Madden: smadden@cea.fr}

\titlerunning{Dissecting the spiral galaxy M\,83}

\authorrunning{A.~Vogler et al.}

\abstract
{We present a detailed mid-infrared study of the nearby, face-on
spiral galaxy M\,83 based on ISOCAM data. M\,83 is a unique case
study, since a wide variety of MIR broad-band filters as well as
spectra, covering the wavelength range of 4 to 18{\m}, were observed
and are presented here.  Emission maxima trace the nuclear and bulge
area, star-formation regions at the end of the bar, as well as the
inner spiral arms. The fainter outer spiral arms and interarm regions
are also evident in the MIR map. Spectral imaging of the
central $3\arcmin \times 3\arcmin$ ($4\kpc \times 4\kpc$) field allows
us to investigate five regions of different environments. The various
MIR components (very small grains, polycyclic aromatic hydrocarbon (PAH)
molecules, ionic lines) are analyzed for different regions throughout
the galaxy. In the total $\lambda$4{\m} to 18{\m} wavelength range,
the PAHs dominate the luminosity, contributing between 60\% in the
nuclear and bulge regions and 90\% in the less active, interarm
regions. Throughout the galaxy, the underlying continuum emission from
the small grains is always a smaller contribution in the total MIR
wavelength regime, peaking in the nuclear and bulge components.  The
implications of using broad-band filters only to characterize the
mid-infrared emission of galaxies, a commonly used ISOCAM observation
mode, are discussed.
We present the first quantitative analysis of new
H$\alpha$ and $\lambda$6~cm VLA+Effelsberg radio continuum maps of
M\,83. The distribution of the MIR emission is compared with that of
the CO, HI, R band, H$\alpha$ and $\lambda$6~cm radio. A striking
correlation is found between the intensities in the two mid-infrared
filter bands and the $\lambda$6~cm radio continuum.
To explain the tight mid-infrared--radio correlation we propose the
anchoring of magnetic field lines in the photoionized shells of gas
clouds.

\keywords{Galaxies: individual: M\,83 -- galaxies: spiral -- galaxies:
ISM -- ISM: dust -- ISM: magnetic fields}

}

\maketitle


\section{Introduction}

Galaxies are composed of stars, gas, dust, dark matter, magnetic fields and cosmic rays,
with each component responding
differently to local and large-scale dynamical effects.
A complete and self-consistent galaxy model
should incorporate information on the stellar distribution and its
relationship to the atomic, molecular and ionized gas and dust and the
subsequent interplay between these ingredients, in order to understand the
physics controlling the heating, cooling and star-formation
processes. These effects will vary widely from galaxy to galaxy on
global scales, as well as within a galaxy between the nucleus, the
bar, spiral arms, interarm regions, and outer disk regions.
We explore various ISM and star formation properties of the well-studied
nearby barred spiral galaxy, M\,83, using ISOCAM high-resolution
mid-infrared (MIR) spectrophotometric and broad-band observations and a high-resolution $\lambda$6~cm radio continuum map
combined from VLA and Effelsberg observations. We compare our data
with results from H$\alpha$, optical, CO, HI and X-ray observations.

M\,83 is the most prominent spiral galaxy in the nearby
Hydra--Centaurus group of galaxies and is one of the largest barred
systems in the sky. Due to its proximity (4.5~Mpc, Thim et al.\
\cite{thim+03}), its near face-on orientation, and relatively low
Galactic foreground absorption ($N_\mathrm{H} \simeq 4\times
10^{20}\cmtwo$, Dickey \& Lockman\
\cite{dickey+lockman90}; E(B-V) = 0.066; Schlegel et al.\ \cite{schlegel+98}), it is ideally suited for the study of
morphological and physical variations of the ISM. M\,83 has been
observed extensively at different energies, ranging from the
long-wavelength radio to the X-ray bands. A rather large data set has
been accumulated including atomic gas (Huchtmeier \&
Bohnenstengel\ \cite{hucht+bohnen81}; Tilanus \& Allen\
\cite{tilanus+allen93}), H$\alpha$ (Ryder et al.\ \cite{ryder+95}; Lundgren et al.\ \cite{lundgren+05}), the molecular gas (Handa et al.\ \cite{handa+90};
Petitpas \& Wilson\ \cite{petit+wilson98}; Crosthwaite et al.\ \cite{crost+02};
Lundgren et al.\ \cite{lundgren+04a}; Lundgren et al.\ \cite{lundgren+04b}; Sakamoto et al.\ \cite{sakamoto+04}) and X-rays (Trinchieri et al.\
\cite{trinchieri+85}; Ohashi et al.\ \cite{ohashi+90}; Okada et al.\
\cite{okada+97}; Ehle et al.\ \cite{ehle+98}; Immler et al.\ \cite{immler+99};
Soria \& Wu\ \cite{soria+wu02}; \cite{soria+wu03}).
M\,83 is also bright in radio continuum due to its
high star-formation rate and strong magnetic field. Highly polarized emission
found in M\,83 suggests a regular magnetic field with a spiral pattern (Neininger
et al.\ \cite{neininger+91}; \cite{neininger+93}).

This study was motivated by the rich assortment of MIR ISO data on M\,83.
By now, almost 10 years after the launch of ISO, we are
beginning to understand the utility of the various tracers in the MIR
wavelength domain.

This paper is organized as follows: Section 2 presents the observational details,
Section 3 the decomposition and description of the
MIR spectral signatures in various selected regions within M83.
The spectral information is then used to decipher the broad-band observations.
This is an important demonstration of what is actually being traced by the
ISOCAM broad-bands; simple interpretations of the broad-bands do not apply everywhere in a
galaxy. We also explore the ratios of the mid-infrared PAH bands
throughout the galaxy. Section 4 presents the radio continuum
interpretation of M\,83. Section 5 presents the MIR observations in
light of H$\alpha$, X-ray emission and molecular gas. In Section 6 we
discuss the spatial distributions of the various tracers and zoom into
the bar, the nuclear region and the south-eastern spiral arm.  In
Section 7, we quantify the correlations between the intensities of the
MIR bands and the other ISM components presented here and discuss the
utility of the MIR characteristics as a tracer of star formation.

\section{Observations and data analysis}

Observation logs and relevant
instrument features for the mid-infrared (MIR) and radio continuum
observations are presented in Table~\ref{isocamdata}
and Sect.~\ref{radio}, respectively. For some of the published images,
we present overlays of our data on the published images
(Figures \ref{lw2-over-radio}--\ref{lw2-over-CO10}, \ref{fig:nuclear_zoom} ,\ref{fig:spiralarm_zoom}). References of the published data used for analysis purposes in this paper, along with instrumental features, are given in Table~\ref{olddata}.

   \begin{table*}
      \caption{     \label{isocamdata}
ISOCAM observations of M\,83. TDT is the Target Dedicated Time Number
of the observation}
             \begin{flushleft}
         \begin{tabular}{lccccccccc}
            \hline
            \noalign{\smallskip}

Filter & Band      & Date & Mean exposure  & Raster    & Field size & TDT\\
       & ($\mu$m)  &      & (sec)  &  mode      & (kpc) & number \\
            \noalign{\smallskip}
            \hline
            \noalign{\smallskip}
LW1 & 4.00--5.00&  16 Aug 1997 & 243 &$2\times9$   & $23\times$7  & 63900451\\
LW2 & 5.00--8.50&  23 Aug 1997 & 224 &$7\times7$   & $17\times$17 &64700246\\
LW3 & 12.0--18.0&  23 Aug 1997 & 224 &$7\times7$   & $17\times$17 &64700350\\
LW4 & 5.50-6.50&   16 Aug 1997 & 243 &$2\times9$   & $23\times$7  & 63900451\\
LW5 & 6.50--7.00&  16 Aug 1997 & 243 &$2\times9$   & $23\times$7  & 63900451\\
LW6 & 7.00--8.50&  08 Aug 1997 & 234 &$2\times9$   & $23\times$7  & 63200147\\
LW7 & 8.50--10.7&  08 Aug 1997 & 234 &$2\times9$   & $23\times$7  & 63200147\\
LW8 & 10.7--12.0&  08 Aug 1997 & 234 &$2\times9$   & $23\times$7  & 63200147\\
LW9 & 14.0--16.0&  08 Aug 1997 & 234 &$2\times9$   & $23\times$7  & 63200147\\
 \noalign{\smallskip}
CVF & 5.00--16.5&  24 Aug 1997 &     &single obs.  & $4\times$4   & 64800263\\
 \noalign{\smallskip}
\hline
         \end{tabular}
         \end{flushleft}
   \end{table*}

   \begin{table*}
      \caption{     \label{olddata}
               References to further M\,83 images used in this study}
             \begin{flushleft}
         \begin{tabular}{llcl}
            \hline
            \noalign{\smallskip}
      & Field size & FWHM  & Reference\\
      & or field of view (FOV) & ($\arcsec$) \\
            \noalign{\smallskip}
            \hline
            \noalign{\smallskip}
Calibrated H$\alpha$ & $12\arcmin\times12\arcmin$ & 1.1 & Lundgren et al.\
   \cite{lundgren+05}\\
\noalign{\smallskip}
%
%
\ion{H}{i} & $\mathrm{FOV}\simeq30\arcmin$  & 12    &  Tilanus \& Allen\   \cite{tilanus+allen93}\\
\noalign{\smallskip}
\noalign{\smallskip}
CO ($J$=1--0)& South-eastern spiral arm & $6.5\times3.5$ & Rand et al.\   \cite{rand+99}\\
CO ($J$=1--0) & Nuclear region and bar$~^1$ & 16 & Handa et al.\   \cite{handa+90}\\
CO ($J$=1--0) & $10\arcmin\times10\arcmin$ & 22 & Lundgren et al.\   \cite{lundgren+04a}\\
CO ($J$=2--1) & $10\arcmin\times10\arcmin$ & 14 & Lundgren et al.\   \cite{lundgren+04a}\\
CO ($J$=3--2) & Nuclear region  & 14 & Petitpas \& Wilson\   \cite{petit+wilson98}\\
\noalign{\smallskip}
X-rays, ROSAT PSPC & $\mathrm{FOV}\simeq2\degr$ & 25--52 & Ehle et
   al.\ \cite{ehle+98}\\
\noalign{\smallskip}
 Radio continuum ($\lambda6$~cm) & $40\arcmin\times40\arcmin$ &
$12 $ & This paper \\
\noalign{\smallskip}
\hline
    \end{tabular}
    \end{flushleft}
$^1$ \ \ And additional points with $7\farcm 5$ stepping close to the nucleus\\
   \end{table*}

\subsection{ISOCAM images}

The MIR data were obtained with the Infrared Space Observatory
(ISO; Kessler et al.\ \cite{kessler+96}) using the ISOCAM
instrument (Cesarsky et al.\ \cite{cesarsky+96a}). This instrument provided imaging
modes with a variety of broad-band filters from $\lambda$4{\m} to 18{\m}
(LW1--LW9) and a spectral imaging mode, using a circular variable filter (CVF)
from $\lambda$5{\m} to 16.5\m. Table~\ref{isocamdata} displays more details of the
various filters used for the data presented here. The ISOCAM detector has a
$32\times32$ pixel field of view, and here one pixel corresponds to
$6\arcsec\times6\arcsec$ ($130\pc \times 130\pc$) on the sky.
The full-width-at-half-maximum (FWHM) of the diameters of the point spread functions (PSF) with a $6\arcsec$ pixel size range from $5\farcs 8$ at 7$\mu$m to $8\arcsec$ for $\lambda$15{\m}.

The LW2 and LW3 broad-band filters, centered at $\lambda$6.7{\m} and
$\lambda$15{\m}, respectively, were used to observe $7\times 7$
raster maps, which contain the full galaxy plus sky
background. The images overlapped by half of an array size, shifted by 1/2 of a pixel, and were
combined to obtain a total image covering a field of $\sim 13\arcmin
\times 13\arcmin$ ($\cor 17\kpc \times 17\kpc$). Other filter bands
were observed in a $2\times9$ raster mode, covering $17\arcmin \times
5.4\arcmin$ ($\cor 23\kpc\times 7\kpc$) fields centered on the nucleus
of M\,83 (see Table~\ref{isocamdata}).  Given the optical extent of the
galaxy of $12\farcm9 \times 11\farcm5$ ($\cor 17\kpc \times 15\kpc$), determination of the field
background is somewhat limited for the $7\times7$ raster maps, while
it is more straight forward for the $2\times9$ raster maps. Thus, we
determined the background from regions of our map outside the optical
extent of M\,83, located between $7\farcm 5$ and 9{\arcmin} from the
center. Subtracting these backgrounds, M\,83 is detected with
a total MIR flux of 20~Jy and 21~Jy in the LW2 ($\lambda$6.75\m) and LW3
($\lambda$15.0\m) band, respectively. Typical errors are of the order of $15\%$,
and our fluxes compare well to those of Roussel et al.\ (\cite{roussel+01b}),
as well as with the
IRAS $12${\m} flux of 26~Jy with an error of the same order as our MIR values.

The filters used for the strip maps include a variety of narrow-bands within the
broader wavelength domain of the LW2 and LW3 bands (cf. Table~\ref{isocamdata} and
the sketched bandwidths in Figs.~\ref{cvfspectra} and \ref{spectraex}).
The narrow-band filters allow us
to interpret the MIR contributions with better accuracy.  While not covering the
complete optical diameter of the galaxy, the strip maps of M\,83 with scan
direction from the north-east to the south-west ($17\farcm 4\times 5\farcm 4$),
offer the advantage of determining the field background more precisely. The
background values for the strip maps were extracted from 2 regions of
$3\arcmin\times 2\arcmin$ (long axis perpendicular to major axis of the strip)
beyond the optical extent of the galaxy. We verified the LW2 and LW3 background
determination from the $7\times7$ raster maps, using the variety of strip maps.

 \begin{figure*}[htb]
 \unitlength1.0cm
 \begin{picture}(18,17.2)
\put(0,8.8){\includegraphics[bb = 68 262 473 619,width=9.36cm,clip=]{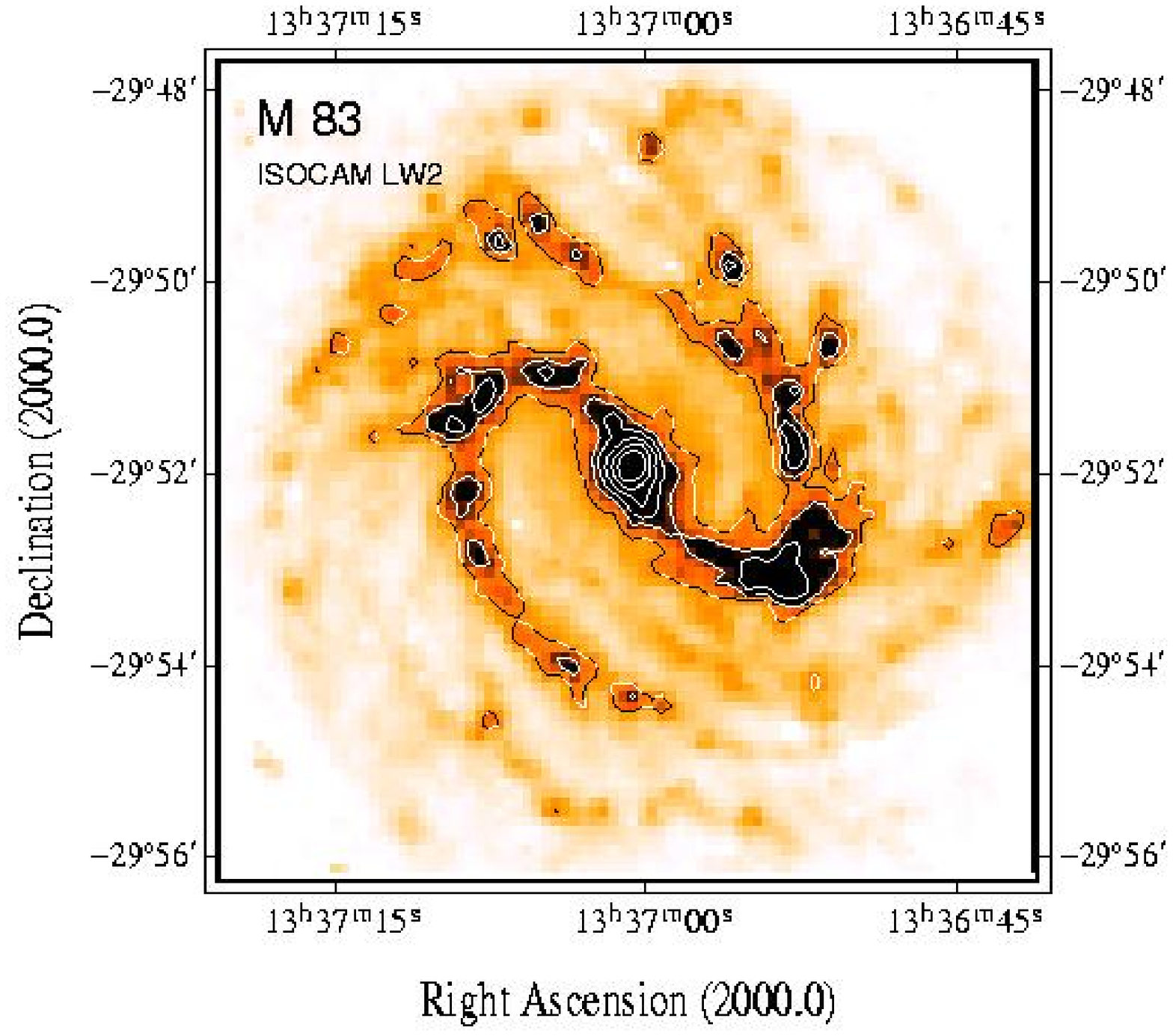}
          }
 \put(9.6,8.8){\includegraphics[bb = 115 262 473 619,width=8.27cm,clip=]{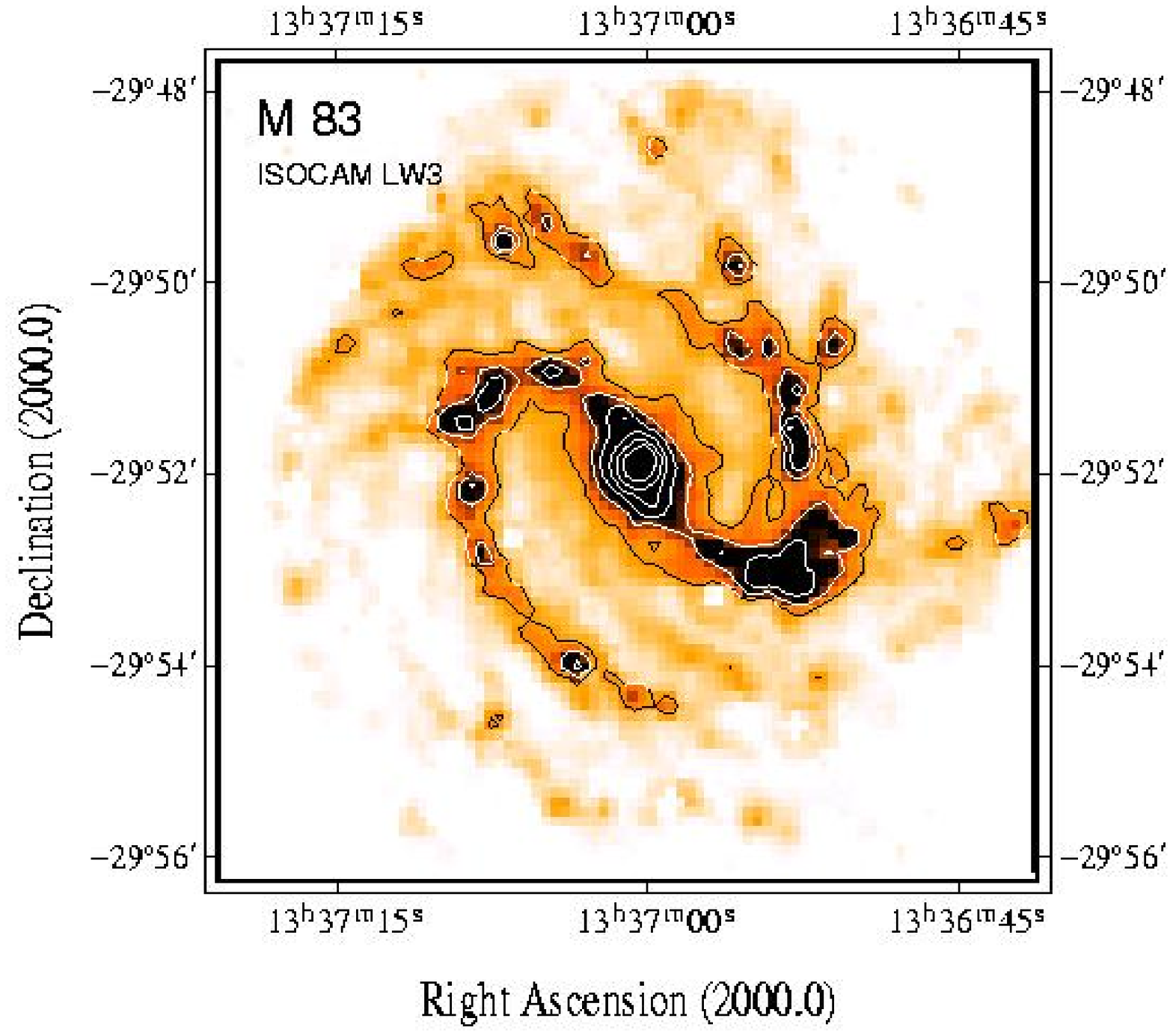}
          }
 \put(0,0){\includegraphics[bb = 68 245 473 619,width=9.36cm,clip=]{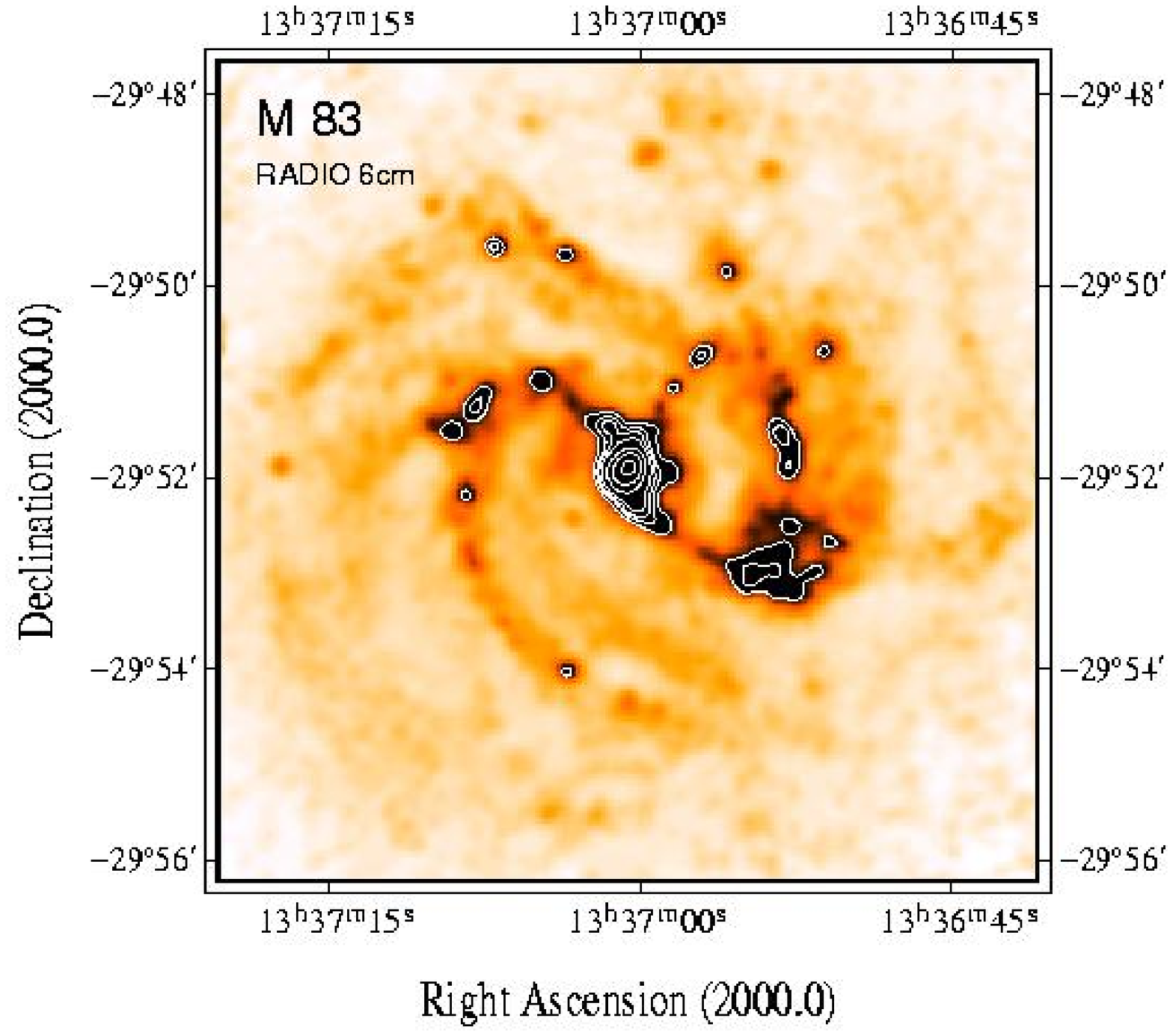}
          }
 \put(9.6,0){\includegraphics[bb = 69 64 500 516,width=8.23cm,clip=]{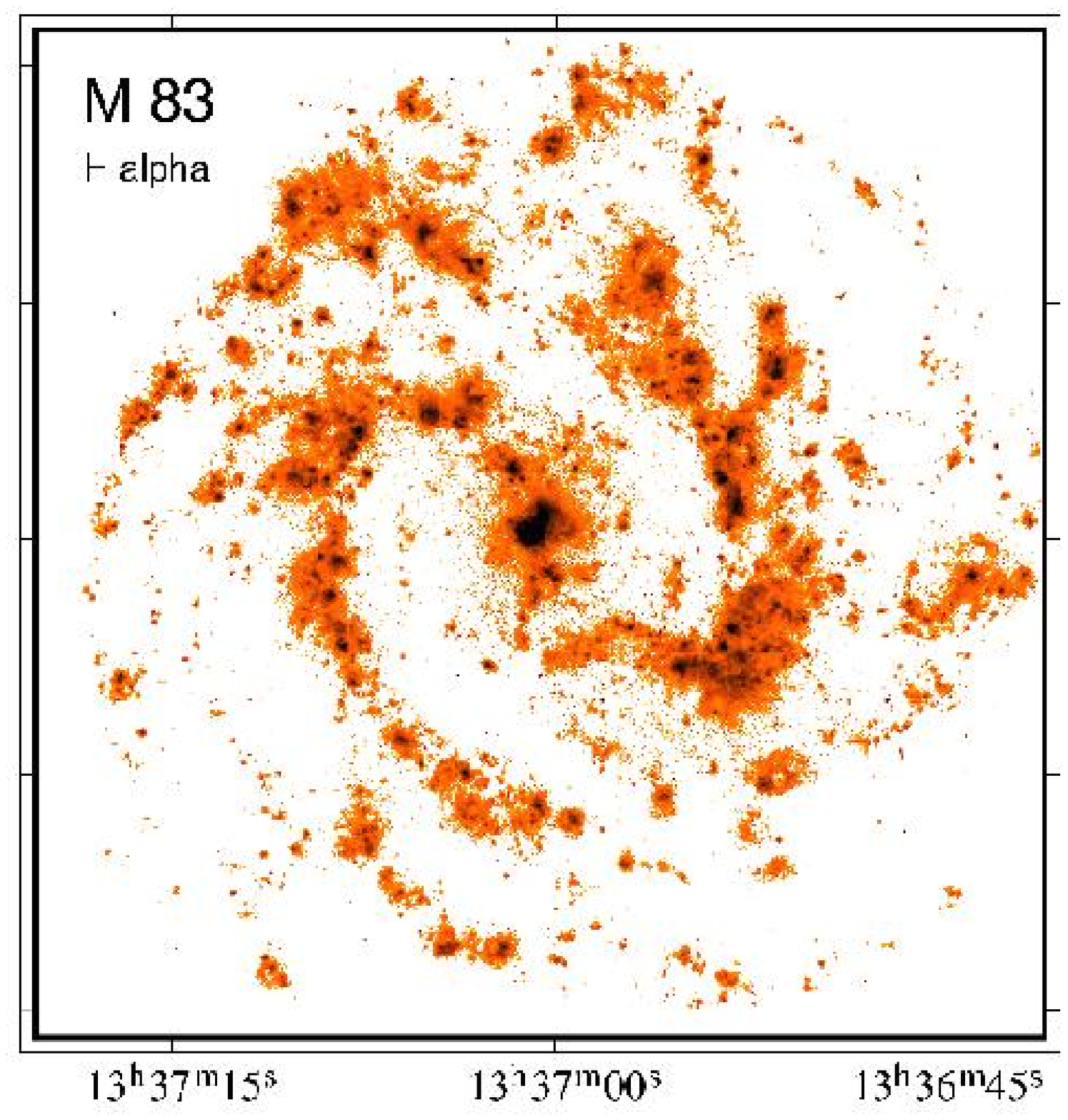}
          }
 \put(8,16){{\huge a)}}
 \put(16.5,16){{\huge b)}}
 \put(8,7.3){{\huge c)}}
 \put(16.5,7.3){{\huge d)}}
 \end{picture}
 \caption{ \label{lw21w3}
 {\it (a)} and {\it (b)}\  ISOCAM LW2 and LW3 images of M\,83 (resolution of $5\farcs 8$ and $8\farcs$, respectively). The dynamical range
of the gray-scale intensity representation runs from 0.5 to $\sim 5$~mJy  per
image pixel (36 arcsec$^{2}$) above the background (8.15~mJy per image
pixel for the LW2 band, 39.3~mJy per image pixel for the LW3 band),
regions of higher intensity have been sketched with contours, namely at
levels of 5, 9, 15, 30, 60 and 120~mJy above the background per image
pixel. \ \
{\it  (c)}\  $\lambda$6~cm combined Effelsberg and VLA radio image.  Contour levels are 2, 3, 4, 5, 10, 30, 50 mJy/beam (angular resolution $12\arcsec$). \ \
{\it (d)}\  H$\alpha$ image of M\,83.
 The continuum-contribution has been subtracted away using
narrow-band data. These observations were carried out with the
Danish 1.54\,m telescope and DFOSC at ESO,  La Silla, Chile.
See Lundgren et al. (\cite{lundgren+05}) for more details.
}
\end{figure*}

The data analysis was performed within the CEA/CIR (Chanial \& Gastaud\
\cite{chanial+gastaud00}) and ESO/MIDAS software environments.

Most cosmic ray impacts, resulting in short duration glitches, were
removed using a multi-resolution median filtering technique in CIR,
while, for the non-classical longer duration glitches, more
sophisticated secondary techniques described in Roussel et
al. (\cite{roussel+01a}) were employed. Transient effects, due to long
term memory effects of the detectors before stabilization is achieved,
were corrected using the Fouks-Schubert method (Coulais \& Abergel\ \cite{coulais+00}).
The flat-field correction was computed by masking the source
and computing the median of the remaining field. See Roussel et
al. (\cite{roussel+01a}) for more specific details.

Figures~\ref{lw21w3}a and b show the entire galaxy in the MIR light
centered at $\lambda$6.75{\m} (LW2; resolution $5\farcs 7$)
and $\lambda$15.0{\m} (LW3; resolution $6\farcs 1$).

Since the broad-band images can contain emission from various dust
components, molecules, ionic species, etc.(see Sect.~\ref{cvf}),
spectro-imaging data were also obtained with a single array setting
of the central $3\farcm2\times3\farcm2$ ($\cor 4\kpc \times
4\kpc$) region of M\,83 to facilitate the interpretation of the broad
band observations. The spectral resolution of the CVF was
~40. Transient correction, deglitching and flat fielding were
performed, basically, as for the broad-band observations described
above. Since the array was never off the galaxy completely,
background subtraction was not done in the usual manner as described
above for photometric mapping observations. Rather, the maximum and
minimum zodiacal light spectra were estimated and subtracted from the galaxy
spectrum, providing limits to the resulting spectrum. This procedure
is described in detail in Roussel et al. (\cite{roussel+01a}).

\subsection{Astrometry corrections of the ISOCAM observations}

To test the position accuracy of the MIR data, we searched
for optical counterparts of the MIR sources outside the optical
extent of M\,83 in the Digitized Sky Survey-2 (DSS-2) image of the M\,83 field
(from the ESO online archive). For the LW2 image, we found counterparts
for five MIR sources (three foreground stars, two background
galaxies). A transformation matrix from the proposed MIR astrometry
to the astrometry implied by the optical image was calculated. A
translation of $3\arcsec$ to the west and $4\arcsec$ to the south,
and, in addition, a clockwise rotation of $1\fdg 4$ was proposed. We
corrected the LW2 image for these values achieving a position
accuracy of $\sim 2\farcs 5$, and then used the LW2 image when
overlaying MIR data on images at other wavelengths.

\begin{figure}
\unitlength=1cm
\includegraphics[bb = 32 27 537 546,width=8.8cm,clip=]{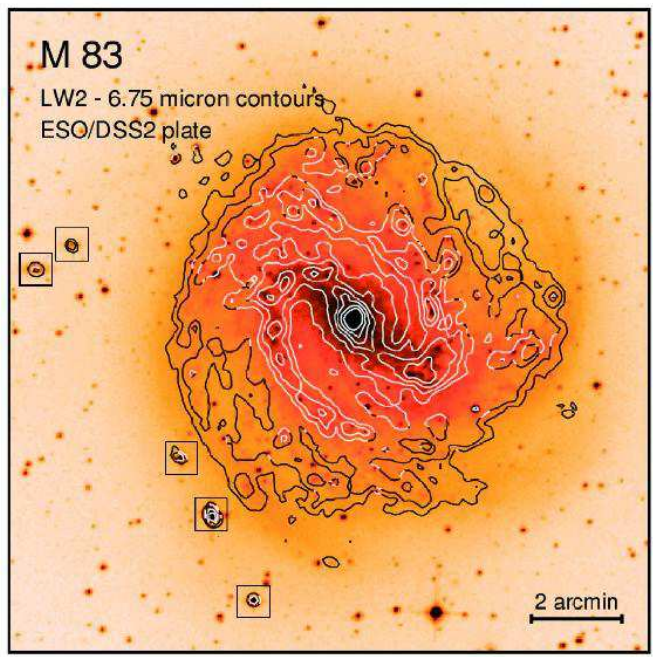}
\caption{\label{final-astro}
ISOCAM LW2 ($\lambda$6.75~$\mu$m central wavelength; angular resolution $6\arcsec$) contours on the DSS-2 image from the ESO archive. Five background galaxies detected outside of M\,83 were used for astrometry and are outlined by boxes. After alignment of the background galaxies with their optical counterparts, the error of the astrometry solution is $\le 2\farcs5$. }
\end{figure}

An overlay of the rotated and shifted LW2 data on the DSS image of M83 is
presented in Fig.~\ref{final-astro}.

\subsection{Radio continuum}
\label{radio}

M\,83 was observed at 4.885~GHz ($\lambda 6.14$~cm) with the VLA synthesis radio
telescopes by NRAO\footnote{The National Radio Astronomy Observatory is a facility
of the National Science Foundation operated under cooperative agreement by
Associated Universities, Inc.} in its DnC configuration on 1992 June 27 and 28 by
S.~Sukumar. To cover most of the galaxy, two fields were observed, centered at
$\mathrm{RA}_{2000}=13^{\rm h} 36^{\rm m} 54^{\rm s},\
\mathrm{DEC}_{2000}=-29\degr 49\arcmin 31\arcsec$ (north-west) and
$\mathrm{RA}_{2000}=13^{\rm h} 37^{\rm m} 09^{\rm s},\
\mathrm{DEC}_{2000}=-29\degr 54\arcmin 06\arcsec$ (south-east). Maps of the two
fields were generated with a synthesized beam of $10\arcsec$ half-power beam
width, corrected for primary-beam attenuation and combined into one map. The rms
noise is $\simeq 20\,\mu$Jy per beam over the region of the main spiral arms, but
increases towards the edges of the map. As the VLA map does not include any
structures of more than about $5\arcmin$ extent, most of the diffuse radio
emission is not detected.

A field of $40\arcmin\times 40\arcmin$ encompassing M\,83 was observed
in 1996 with the Effelsberg 100-m telescope\footnote{The 100-m
Effelsberg telescope is operated by the Max-Planck-Institut f\"ur
Radioastronomie on behalf of the Max-Planck-Gesellschaft (MPG).}
using the 4.85~GHz receiver system providing an angular resolution of
$2\farcm 6$. The final map has an rms noise of $\simeq 1$~mJy per
beam.  Emission from M\,83 is detected out to about $6\arcmin$
($\cor$ 8~kpc) radius.

The VLA and Effelsberg data were merged in the Fourier plane using
IMERG based on NRAO's SDE software (see Beck \& Hoernes\
\cite{beck+hoernes96}). The final map has the same angular resolution
and rms noise as the original VLA map, but includes all diffuse extended
emission (about 35$\%$ of the total flux) detected with the Effelsberg
single-dish. The map is shown in Fig.~\ref{lw21w3}c. The polarization
map was presented by Beck (\cite{beck02}) and will be discussed in a
future paper (Beck et al.\ \cite{beck+05}).

\section{Results}

\subsection{The ISOCAM MIR View of M\,83}
\label{cvf}

The ISOCAM CVF wavelengths ranges from $\lambda$5{\m} to 17{\m} and
traces a variety of dust components as well as ionic and molecular
gas lines, depending on local environmental conditions within galaxies.
The MIR dust components primarily consist of 2 types in galaxies: (1)
small particles or large molecules (e.g.,
polycyclic aromatic hydrocarbon molecules; PAHs
\footnote{Although the precise nature of these particles is uncertain,
we will refer to these as PAH bands throughout the paper.}),
stochastically heated,
undergoing large temperature fluctuations of the order of hundreds of K on short
timescales as a result of single photon heating processes
(e.g. Puget \& Leger\ \cite{puget+84}; Sellgren et al.\
\cite{sellgren+90}; Allamandola et al.\ \cite{allamandola+89};
Verstraete et al.\ \cite{verstraete+01}; Kr\"ugel\ \cite{krugel03}),
with major emission bands primarily at $\lambda$3.3\m, 6.2\m, 7.7\m,
8.6\m, 11.3{\m} and 12.7{\m}; (2) very small grains (VSGs) which
typically have sizes in our Galaxy in the range of 1~nm--150~nm
(D\'esert et al.\ \cite{desert+90}) and which are the origin of the
MIR continuum in our ISOCAM spectra, longward of $\lambda$11 \m. These
VSGs, representing a range of temperatures, could be either
stochastically heated or in thermal equilibrium, depending on the
radiation field properties and grain sizes. Both PAHs and VSGs are
excited mostly via UV photons in the vicinity of active star formation
regions. The PAHs are observed to peak in the photodissociation
regions (PDRs), which are the interface regions between molecular
cores and \ion{H}{ii} regions. The VSGs emit prominently in the
nebular regions (e.g. Cesarsky et al.\ \cite{cesarsky+96b}; Verstraete
et al.\ \cite{verstraete+96}).  On the other hand, PAHs can also be
excited by less energetic means, such as optical photons, as evidenced
by their presence in the diffuse interstellar medium (ISM) (Chan et
al.\ \cite{chan+01}) and elliptical galaxies (e.g. Athey et al.\
\cite{athey+02}; Xilouris et al.\ \cite{xilouris+04}).

In the case of more active regions, such as the nuclei of normal
galaxies, starbursts or active galactic nuclei, various nebular lines
can also be observed in the ISO MIR wavelength range (e.g. Genzel et
al.\ \cite{genzel+98}; Rigopoulou et al.\ \cite{rigopoulou+02}; Sturm
et al.\ \cite{sturm+02}; Lutz et al.\ \cite{lutz+03}; F\"orster
Schreiber et al.\ \cite{forster+03}; Verma et al.\ \cite{verma+03};
Madden et al.\ \cite{madden+05}). In spiral galaxies
(i.e. Roussel et al.\ \cite{roussel+01a};
\cite{roussel+01b}) the nuclear region can show a $\lambda$12.8{\m} [NeII] line,
and in harder radiation fields the $\lambda$15.6{\m} [NeIII] line can be present, too.
A more detailed discussion of the general MIR characteristics of spiral galaxies
and starbursts is given, e.g., in Vigroux et al. (\cite{vigroux+99}), Dale et al.
(\cite{dale+00}), Genzel \& Cesarsky (\cite{genzel+cesar00}) and Kr\"ugel (\cite{krugel03}).

\subsection{Spectral properties of the nucleus, spiral arms, bulge and interarm regions}

{\footnotesize
\begin{table*}
\caption{   \label{cvf-spectra-tab}
Comparison of MIR components using CVF observations of the central $3\arcmin$ field of M\,83. (Note: 'Cont.' in the table refers to the continuum emission.)}
\begin{flushleft}
\begin{tabular}{lcccccccccc}
\hline
\noalign{\smallskip}
Region&PAHs&Cont.& PAHs:&PAHs:&PAHs:&PAHs:&LW2/LW3&LW9/LW6& PAHs  & PAHs \\
(see& $10^{-11}$ & $10^{-11}$& total &Cont. &Cont. & Cont.  & from CVF    & from CVF &
         6.2 : 7.7 : 8.6 : 11.3 & 11.3/12.8\\
Fig.\ref{cvfspectra}) & erg/s& erg/s& 4--18& 4--18& LW2 & LW3$^{1}$ & 6.7/15$\mu$m & 15/7.7$\mu$m & $\mu$m & $\mu$m \\
&/cm$^2$&/cm$^2$& $\mu$m & $\mu$m \\
\noalign{\smallskip}
\hline
\noalign{\smallskip}
{\it (1) \  } & 10.7 &6.6& 0.62 & 1.6& 13 & 0.26 & 0.57 & 1.1 &0.32 : 1 : 0.16 : 0.22& 1.25\\
{\it (2) \ } & 71.3 & 31.5 & 0.69 &2.3 &17 & 0.34  & 0.69 & 0.8 &
0.31 : 1 : 0.22 : 0.29&1.49\\
{\it (3) \  } & 42.0 & 17.7 & 0.70 &2.4&  13 & 0.6 & 0.91 & 0.6
&0.37 : 1 : 0.19 : 0.33 & 1.37\\
{\it (4) \ } & 3.4 & 1.1& 0.76 & 3.0& 20  & 0.7 & 1.22 &  0.4
&0.29 : 1 : 0.26 : 0.29 & 1.41\\
{\it (5) \  } & 2.7 & 0.3 & 0.90 & 9.1& $\ge 100$ &3.3 & 1.75 & 0.2
&0.42 : 1 : 0.11 : 0.59 &2.48\\
\noalign{\smallskip}
\hline
\end{tabular}
\end{flushleft}
$^{1}$ A possible blend of the PAH band with the
[Ne{\sc ii}] line at $\simeq\lambda$12.8{\m}
cannot be resolved by the CVF. Typical contributions to the [Ne{\sc ii}] line
are expected to
be small (e.g. Roussel et al.\ \cite{roussel+01a}; Sturm et al.\ \cite{sturm+00}),
and are assumed to be below 30$\%$, with the remainder attributed to the
PAH feature.
\end{table*}
}

\begin{figure}
\unitlength=1cm
\begin{picture}(9,17)
\put(2.65,12.3){\includegraphics[bb = 47 156 522 631,width=4.9cm,clip=]{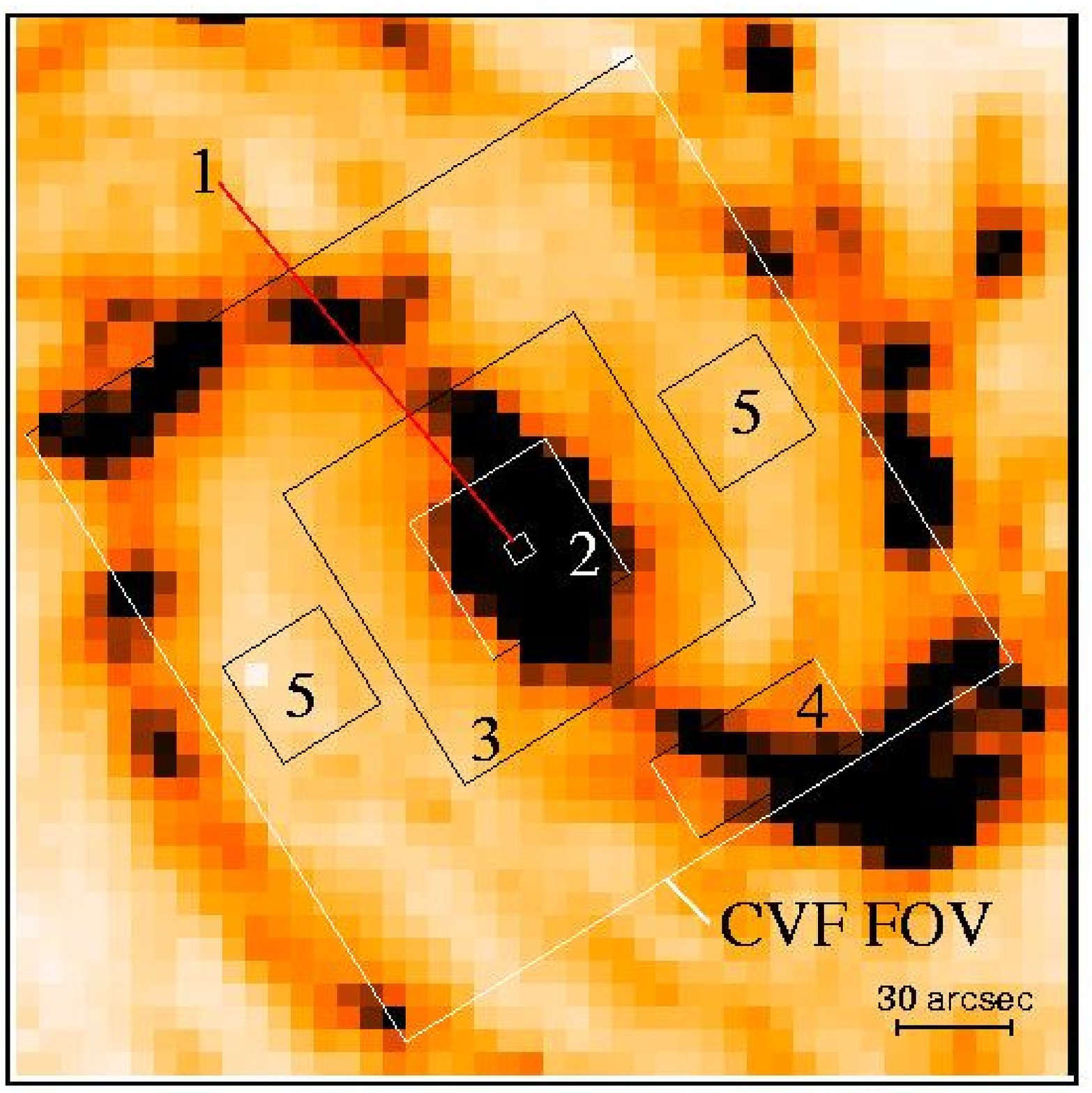} }
\put(0,0){\includegraphics[bb = 54 360 558 730,width=8.9cm,clip=]{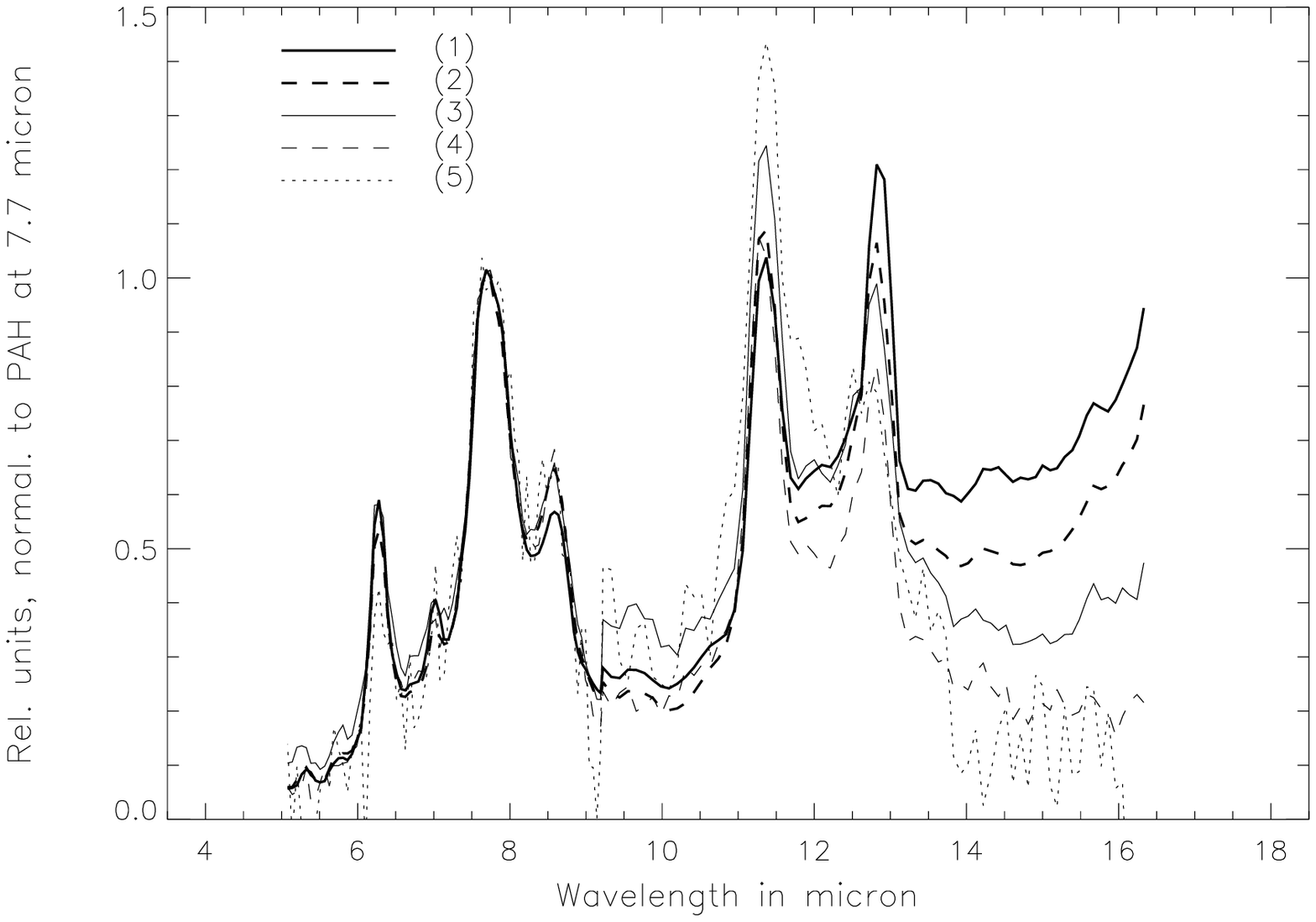} }
\put(0,6.5){\includegraphics[bb = 54 360 558 720,width=8.9cm,clip=]{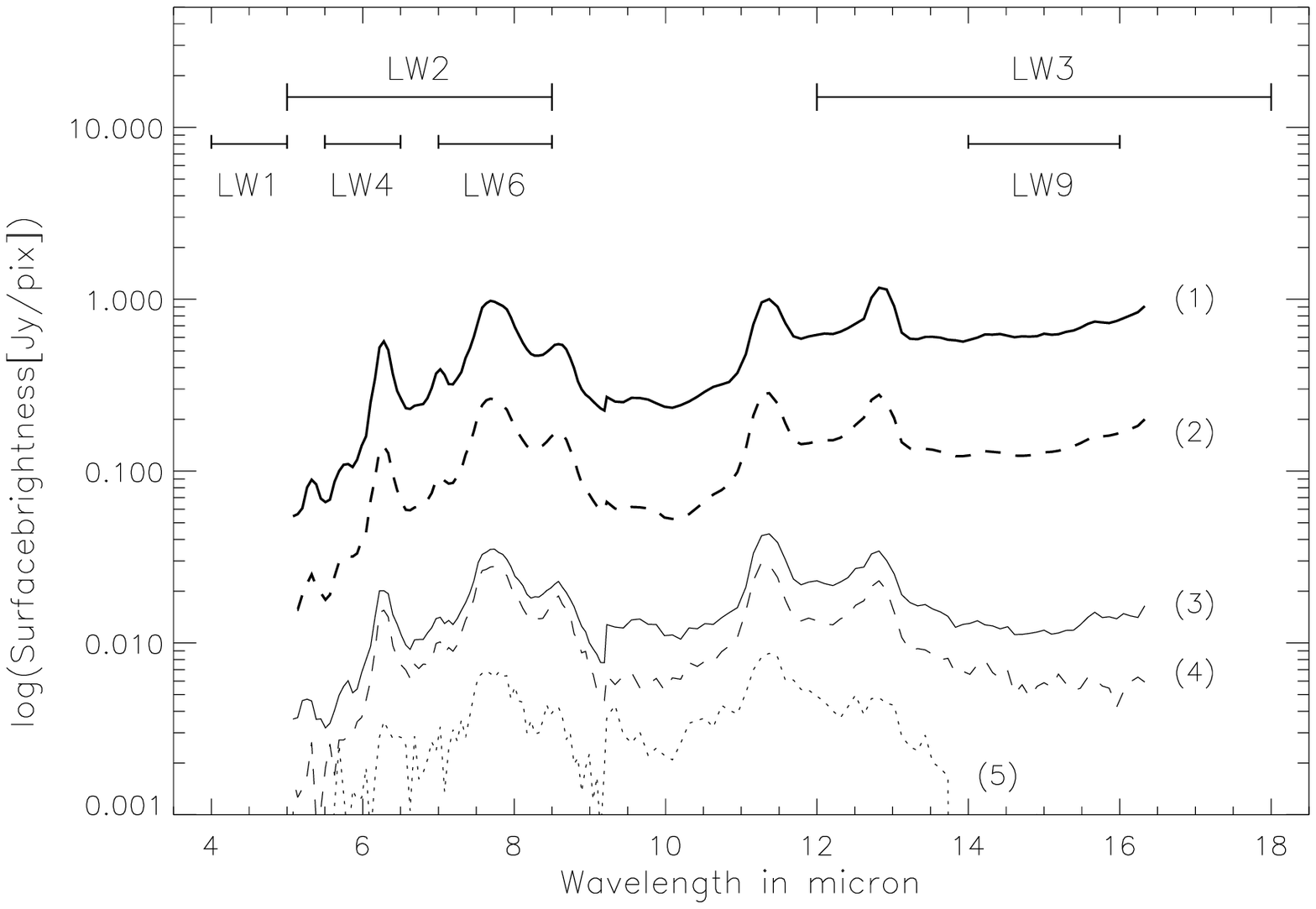} }
\end{picture}
\caption{ \label{cvfspectra}
{\it Top panel:} The boxes mark the ISOCAM CVF field of view and
extracted regions for spectra on top of the ISOCAM LW2 ($\lambda$6.75{\m}
central wavelength) image. \ \
{\it Middle panel:} \label{cvfima}\label{schoenima}
CVF spectra for five different regions, namely (1) the central pixel of
the M\,83 field; (2) the $7\times7$ central pixels excluding the region
(1); (3) the central $11\times11$ pixels excluding regions (1) and (2);
(4) a $9\times4$ pixel on-arm region offset from the bulge;
(5) average of two $5\times5$ pixel interarm regions. \ \
{\it Bottom panel:}  The same spectra normalized to the
$\lambda$7.7~$\mu$m PAH emission and on a linear scale.
}
\end{figure}

\begin{figure}
\unitlength1cm
\begin{picture}(8.9,15)
\put(.4,2.6){\includegraphics[bb = 96 395 541 701,width=8.55cm,clip=]{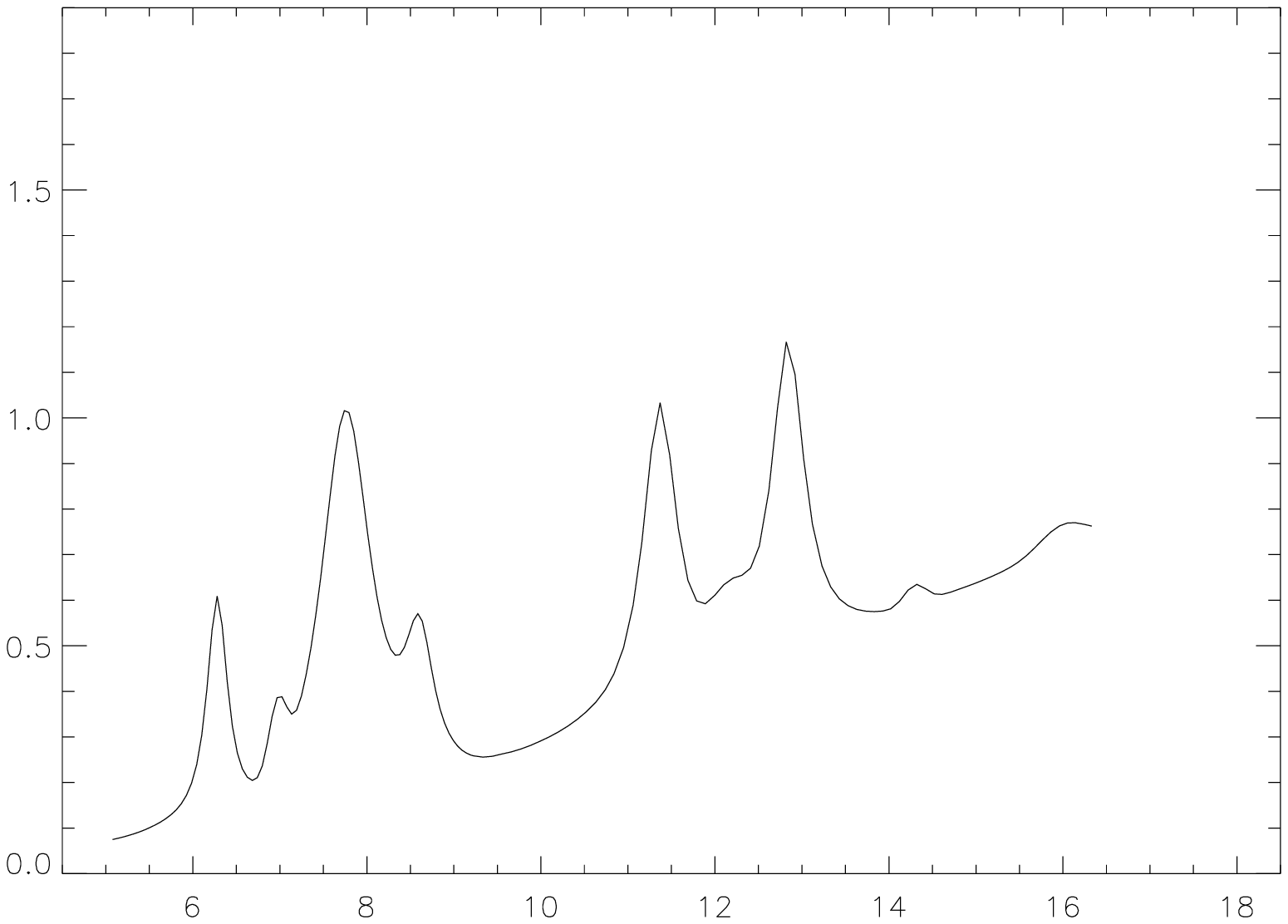} }
\put(.4,8.6){\includegraphics[bb = 96 395 541 701,width=8.55cm,clip=]{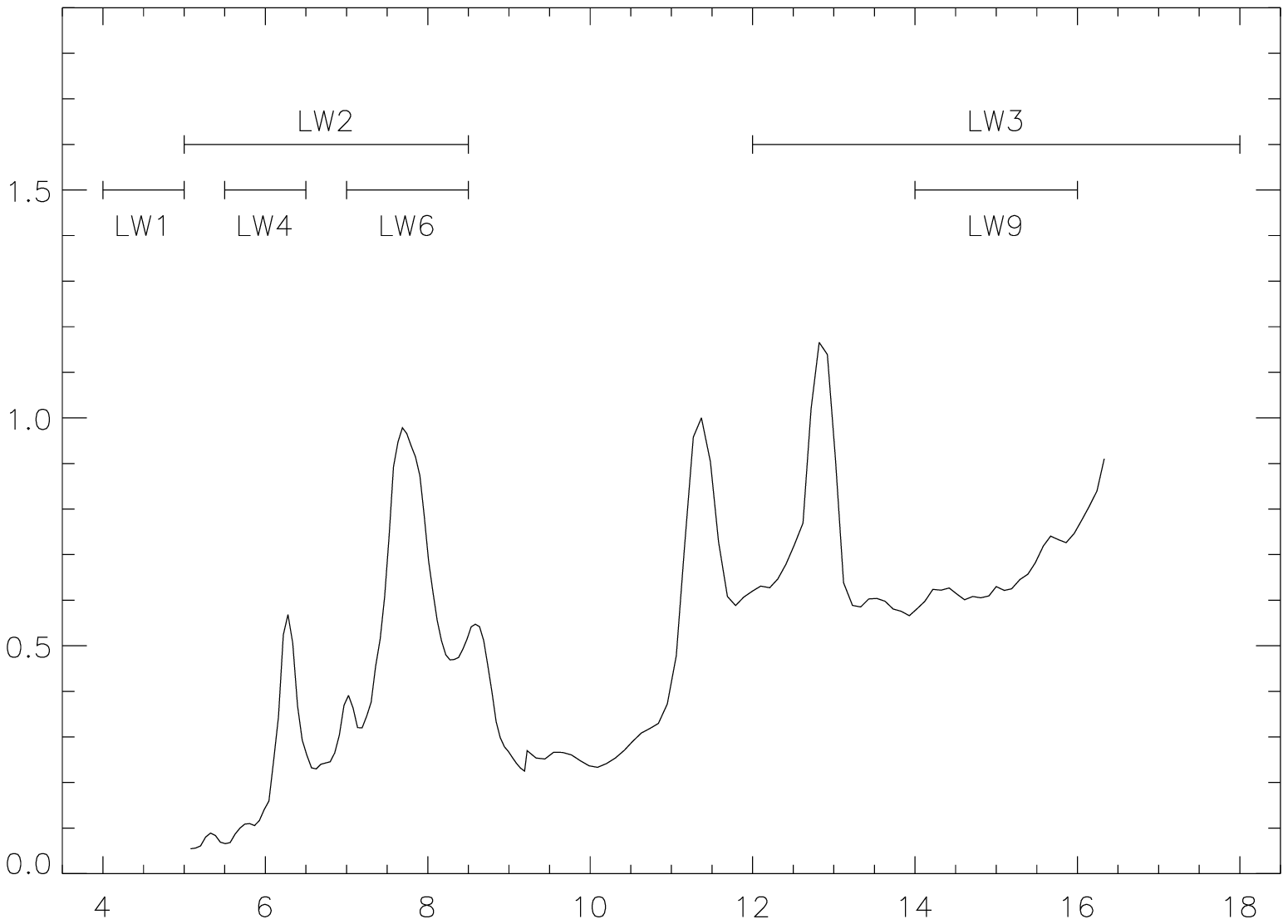} }
\put(.4,0.6){\includegraphics[bb = 69 380 510 470,width=8.55cm,clip=]{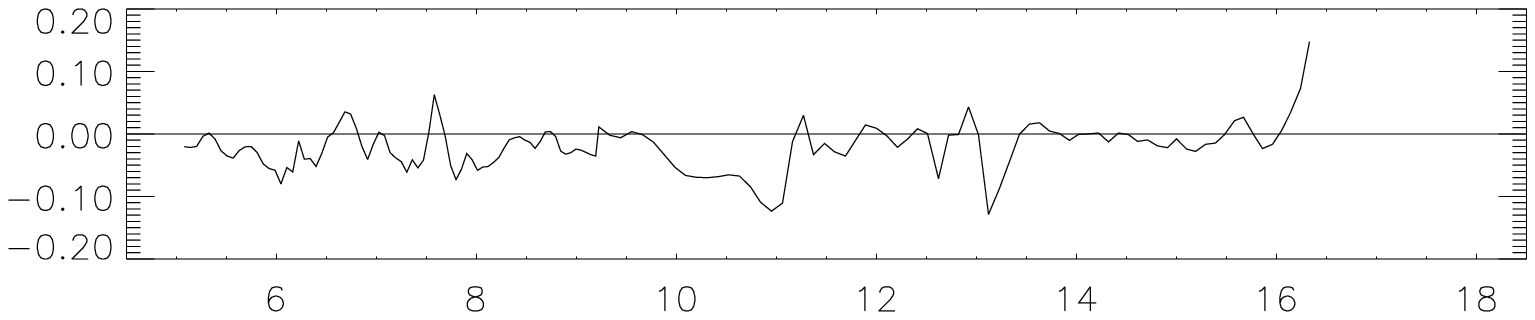} }
\put(4.4,0.0){$\lambda$ ($\mu$m)}
\put(0,10.5){\rotatebox{90}{$f_{\nu}$ (Jy)}}
\put(0,4.5){\rotatebox{90}{$f_{\nu}$ (Jy)}}
\end{picture}
\caption{ISOCAM MIR spectrum of the central pixel of the M\,83 observation.
The x-axis gives the wavelength in {\m}, the y-axis the flux in Jy at
the spectral resolution of the ISOCAM CVF. \ \ {\it Top panel:}
Observed spectrum. The 5 PAH bands figure
prominently at 6.2{\m}, 7.7{\m}, 8.6{\m}, 11.3{\m} and
12.7{\m}, while the ionic line ([Ar{\sc ii}] is also seen at
$\lambda$7.0{\m}.
The bandwidths of the LW1, LW2, LW3, LW4, LW6 and
LW9 filters are indicated.\ \ {\it Middle panel:} Fitted lines plus continuum.
The spectra of all the fitted lines and the continuum were
co-added and represented as the modeled spectra. The model details are
described in section~\ref{regionsect}.
\ \ {\it Bottom panel:}  Difference between observed and fitted
spectrum.
}
\label{spectraex}
\end{figure}

In order to determine
the MIR properties of various environments within a typical spiral galaxy, such
as M\,83, the galaxy is subdivided into the following regions (outlined
in Fig.~\ref{cvfima}) from which we extracted spectra for comparison:

\label{regionsect}

\begin{enumerate}
\item The central and brightest $6\arcsec\times 6\arcsec$
      ($130\pc \times 130\pc$) pixel around the nucleus of M\,83. This
      includes much of the starburst arc of the central region of the galaxy
	(Sakamoto et al.\ \cite{sakamoto+04})

\item A $42\arcsec\times 42\arcsec$ ($910\pc\times 910\pc$) region
      centered on the nucleus, excluding the innermost pixel and
      containing the extended circumnuclear environment

\item A $90\arcsec\times 90\arcsec$ ($2\kpc\times 2\kpc$) region
      highlighting the bulge and centered on the nucleus, excluding the nuclear
      regions (1) and (2)

\item A $24\arcsec\times 54\arcsec$ ($520\pc \times
      120\pc$) region centered on the SW spiral arm and excluding the
      bulge region

\item An interarm region containing two $30\arcsec\times 30\arcsec$
      ($650\pc \times 650\pc$) regions between the bulge and inner
      spiral arms.
\end{enumerate}

Figure~\ref{cvfspectra} ({\it middle panel}\/) displays the variety of
spectra for the different regions.  For the spectrum of region (5), the
interarm region, the 2 regions on either side of the bulge are
averaged to enhance the signal to noise ratio. The most striking
features identified in the spectra of Fig.~\ref{cvfspectra} toward all
regions are the PAH bands at $\lambda$6.2{\m}, 7.7{\m}, 8.6{\m} and
11.3{\m}, and a blended [Ne{\sc II}] and PAH feature at
$\simeq\lambda$12.8{\m}.  There is a general tendency for the
continuum emission from hot grains beyond $\simeq\lambda$13{\m} to
peak toward the nucleus, and to decrease toward the outer, more
quiescent regions, as demonstrated in Fig.~\ref{cvfspectra} ({\it
bottom panel}\/, where the spectra are normalized at
$\lambda$7.7{\m}). This was also illustrated for other spiral galaxies
in Roussel et al.\ (\cite{roussel+01a}) and starburst galaxies
(i.e. F\"orster-Schreiber et al.\ \cite{forster+03}). The shape and
the intensity of the spectra from $\lambda$5{\m} to 11{\m}, dominated
by the cluster of PAH bands, varies less so throughout the galaxy.

For quantitative analysis we decomposed the spectra, fitting them with 3
components (Laurent et al.\ \cite{laurent+00}): 1) PAH bands (the major PAH bands:
$\lambda$3.3{\m}, 6.2{\m}, 7.7{\m}, 8.6{\m}, 11.3{\m} and 12.7{\m} and the
less prominent bands: $\lambda$5.3{\m}, 5.7{\m}, 12.0{\m}, 13.6{\m} and 14.3{\m})
were fitted with Lorentzian profiles (i.e. Boulanger
et al.\ \cite{boulanger+98});
2) the ionic lines ([Ar{\sc ii}] $\lambda$7.0{\m}, [Ar{\sc iii}] $\lambda$9.6{\m},
[S{\sc iv}] $\lambda$10.5{\m}, [Ne{\sc ii}] $\lambda$12.8{\m} and [Ne{\sc iii}]
$\lambda$15.6{\m}) were modeled with Gaussian profile characteristics and 3) the
continua were assumed to have power-law shapes. The central wavelengths
were fixed and the program computed the width and intensity. The index of the
continuum power law was a free parameter. The different
spectral components of the CVF spectra are illustrated in
Fig.~\ref{spectraex} for the case of region (1). The figure shows the
observed CVF spectrum (above panel), the model lines and bands plus continuum
for comparison (middle panel). The very
low residual emission (Fig.~\ref{spectraex} bottom panel) after the
subtraction of the model spectrum from the observation, speaks well
for the assumptions in the modeling of the CVF spectra. The model
fits are used for quantitative comparisons in the following discussion.

Typical flux densities for the brightest PAH features are $\ga
0.5$~mJy, while atomic line contributions are smaller, the largest
being the [Ne{\sc ii}] $\lambda$12.8\m\ line (0.2~mJy) towards the nucleus. We detect
minor contributions from the [Ar{\sc ii}] 7.0{\m} line.
The [Ne{\sc iii}] 15.6{\m} line appears to be present in the nucleus
(regions 1 and 2) and the bulge (region 3) and not evident in regions 4 (on arm) and 5
(interarm). Throughout the galaxy, PAHs seem to be ubiquitous and
account for the majority of the MIR 4 to 18{\m} luminosity
(Table~\ref{cvf-spectra-tab}). The continuum emission is at most 40\%
of the total MIR emission toward the central starburst core (region
1), and decreases away from the central region. Toward the interarm
region (region 5), the PAHs account for almost all of the MIR energy.
Both the LW2 and LW3 images (Sect.~\ref{lw2lw3ref}), as well as the
CVF spectra, show MIR emission in the form of PAH bands but no
significant continuum towards the interarm or outer regions.
Toward diffuse regions of the Galaxy, the PAH bands are {\it the}
fundamental component of the MIR emission (e.g. Mattila et al.\
\cite{mattila+96}; Lemke et al.\ \cite{lemke+98};
Chan et al.\ \cite{chan+01}).

\subsection{Using the CVF to determine the broad-band composition}

The ISOCAM LW2 ($\lambda$5.0--8.5{\m}) and LW3
($\lambda$12.0--18.0{\m}) bands were popular broad-bands observed with
ISO and often were the only observations used to characterize the
MIR ISM of galaxies. Depending on the type of galaxy, these bands
encompass various components to varying degrees. Often it was simply
assumed that the LW2 band is dominated by PAH bands, while the LW3
traces the dust continuum. Here we take advantage of the fact that a
variety of broad-bands as well as CVF spectra were obtained for M\,83
to quantify the contribution of the various MIR components to each of
these broad-bands, using the CVF spectra. We can then interpret the
ISOCAM MIR broad-band LW2/LW3 ratio in galaxies more accurately.

In Table~\ref{cvf-spectra-tab} we present the results for the modeled
CVF for the different regions in M\,83 (refer to Fig.~\ref{cvfspectra}
for definitions of the region numbers) extracting continua and PAH
band contributions. The cluster of PAH bands at $\lambda$ 8{\m}
($\lambda$6.2, 7.7, 8.6{\m}) is always dominant in the LW2 broad-band
filter throughout M\,83, with the continua being on the order of about
5\% of the total LW2 broad-band flux. We do not detect any substantial
contribution from the very hot grain continua or stellar emission from
evolved populations, emitting below $\lambda$9{\m} in the CVF
spectra. Hot grain continuum emission at these shorter MIR wavelengths
has been detected in active galactic nuclei and prominent starbursts
(Hunt et al.\ \cite{hunt+02}; Madden et al.\ \cite{madden+05}). In
early-type galaxies, this spectral range is dominated by stellar
emission (e.g. Xilouris et al.\ \cite{xilouris+04}). A substantial
fraction of the flux in the LW3 band arises from the continuum emission:
75\% to 80\% in the central region, $\simeq$60\% in the bulge and spiral
arms. In the more quiescent interarm regions, however, the $\lambda$12.7\m\ PAH band
dominates the LW3 band emission ($\simeq$75\%).

While it is tempting to use the LW2/LW3 ratio to study the MIR emission
in galaxies, given the different emission features contributing to the
broad-band filters, it would be preferable to choose a filter isolating
a PAH band only, and another which selects only the continuum of
$\lambda>10\,\mu$m. The LW6 broad-band filter ($\lambda$7.0{\m} to
8.5{\m}), for example, only traces the $\lambda$7.7{\m} PAH band and part of the $\lambda$8.6{\m} PAH band while
the LW9 ($\lambda$14.0{\m} to $\lambda$16.0{\m}) filter traces the
$\lambda$15{\m} continuum (when the [Ne~III] line is not present, as is
the case here), removing possible ambiguities
(Fig.~\ref{spectraex}). The CVF predicts a ratio of LW9/LW6
($\lambda$15{\m} continuum) / ($\lambda$7.7{\m} PAH) of 1.1 for the
nuclear region and decreases to 0.2 for the diffuse interarm region. In M\,83 the
dynamical range of the LW9/LW6 intensity ratio (factor of 5, see
Fig.~\ref{lw9-over-lw6}) is larger than that of the LW2/LW3 ratio
(factor of 2, see Sect.~\ref{lw2lw3ref}), making it a more sensitive tool to distinguish different physical regions of a galaxy.

\subsection{PAH band ratios in M\,83}

The ratios of the PAH band intensities are a means of testing the
validity of the PAH model assumption for the $\lambda$6.2{\m},
7.7{\m}, 8.6{\m} and 11.3{\m} and 12.7{\m} bands (e.g. Jourdain de
Muizon et al.\ \cite{jourdain+90}; Schutte et al.\ \cite{schutte+93};
Allamandola et al.\ \cite{allamandola+99}).
We normalize each spectrum to the prominent $\lambda$7.7{\m} PAH band
(Fig.~\ref{cvfspectra}) and inspect the PAH band ratios in the
different regions in M\,83. The ratios of the PAH bands 6.2{\m} :
7.7{\m} : 8.6{\m}: 11.3{\m} vary, at most, by a factor of 1.5 for regions
1, 2, 3 and 4 (Table~\ref{cvf-spectra-tab}), which sample very
different ISM conditions from the starburst core, to a spiral arm and
bulge.  The PAH ratios, however, deviate more significantly in the
diffuse, interarm region (region 5), with a particularly large value
of 0.59 for the 11.3/7.7 ratio, $\sim$ 2.5 times larger than toward the nuclear region. Large variations of the 11.3/7.7
ratio have been seen in a wide variety of Galactic and extragalactic
sources (Vermeij et al.\ \cite{vermeij+02}; Galliano\
\cite{galliano04}) with no obvious systematic variation with radiation
field intensity or hardness (Chan et al.\ \cite{chan+01}). PAH band
ratios vary over only 40\% in reflection nebulae with 3 orders of
magnitude variation in radiation field intensity (Uchida et al.\
\cite{uchida+00}).  A precise conclusion as
to what is controlling these observed variations is not yet
clear. Differences in chemical structure, such as compact versus open
structures, may be one explanation for the observed variations
(Vermeij et al.\ \cite{vermeij+02}). Laboratory experiments may
eventually shed light on the nature of the observed PAH band ratios.

\label{beforesection}

\subsection{LW2 and LW3 broad-band images of M\,83}
\label{lw2lw3ref}

Fig.~\ref{final-astro} shows highly structured MIR emission from the
entire optical extent of M\,83. The brightest MIR region is coincident
with the nucleus of M\,83. The emission from the bulge region is
extended, at the ISOCAM resolution, along the leading edge of the bar
of the galaxy. The south-western and north-eastern ends of the bar are
associated with prominent MIR emission regions, which are also present
in the radio and in H$\alpha$ maps (Sec.~\ref{radioinf}). The MIR
morphology argues for the existence of gas and dust in these
bar/interstellar medium interfaces.

The inner spiral arm pattern of the galaxy is prominent in the LW2 and
LW3 bands (Fig.~\ref{lw21w3}a and b) while MIR diffuse emission of
lower intensity is detected from the interarm regions. Both LW2 and
LW3 images show a lane of enhanced emission in the interarm region
between the bulge and the south-eastern inner spiral arm. The
``interarm lane'' is also visible on the radio image
(Fig.~\ref{lw21w3}c) as well as on the optical image
(Fig.~\ref{final-astro}). It is a narrow spiral arm (highly polarized
in radio continuum, see Fig.~3 in Beck\ \cite{beck02}) with a
relatively low star-formation rate.

The morphology of the galaxy is very similar in LW2 and LW3, hence, we
have chosen the LW2 map for overlays on images of other wavelengths,
since the signal-to-noise ratio of the LW2 map is superior to that of
the LW3 map.

The ratio of LW2/LW3 is known to vary little across the disk of a
spiral galaxy and observed to be of the order of 1 in most galaxies
(Helou et al.\ \cite{helou+96}; Vigroux et al\ \cite{vigroux+99};
Roussel et al.\ \cite{roussel+01a}; Roussel et al.\
\cite{roussel+01b}). These MIR bands have often been used as a
diagnostic along with FIR colors in an ISO-IRAS color-color diagram
(Vigroux et al.\ \cite{vigroux+99}; Dale et al.\
\cite{dale+00}). LW2/LW3 $\sim$ 1 when the LW2 band is composed
completely of PAH bands and the LW3 band has both continuum and PAH
band emission, in mostly equal parts. This occurs when the ISM is
dominated by neutral regions and photodissociation regions (PDRs), as
is the case for much of the disk material in galaxies. Toward \ion{H}{ii}
regions, the rapidly-rising hot dust continuum emission accounts for most of
the LW3 band, thus, the LW2/LW3 ratio decreases to $\leq$ 0.5. This
ratio does not vary significantly from normal, quiescent galaxies to
mildly-active galaxies as the IRAS f(60$\mu$m)/f(100$\mu$m) increases.
However, starburst galaxies and AGNs show a dramatic decrease in the
LW2/LW3 ratio due to two effects operating simultaneously,
the suppression of the PAH band emission in LW2,
presumably through the destruction of its carriers,
and the dominance of
hot 15$\mu$m dust continuum emission in LW3 (e.g. Dale et al.\
\cite{dale+00}; Roussel et al.\ \cite{roussel+01b}).
Toward the starburst core and surrounding nuclear region of M\,83
(regions 1 and 2) the LW2/LW3 ratios are between 0.57 and 0.69
(Table~\ref{cvf-spectra-tab}), where the continuum emission accounts for 75
to 80\% of the total MIR emission. The ratio for the bulge (region 3)
is about 90\%, while in regions 4 and 5 the PAH bands contribute 50\%
or more to the total MIR emission.

M\,83 is a good illustration of the diagnostic capability of the MIR
colors. Using the CVF spectra to quantify the components of the
LW2 and LW3 broad-bands, we see that the LW2/LW3 ratio does provide
an accurate tool to probe the PAH band versus hot dust contributions
in the more active regions, such as the nucleus (see also Roussel et
al. \cite{roussel+01a, roussel+01b}). However, this broad-band ratio
becomes more ambiguous outside of the nuclear region, as the PAH band
to continuum ratio increases. In these regions, it is necessary to
have other accurate indicators of the MIR components contained within
these broad-bands.

\begin{table*}
\caption{ Measured fluxes $f_{\nu}$ and luminosities $L$ ($D =
          4.5$~Mpc) in the different MIR filters }
\label{lwfluxes}
\begin{flushleft}
\begin{tabular}{lccccccc}
\hline
\noalign{\smallskip}
&
Central $\lambda$ &
$f_{\nu}(\mathrm{d\le1'})$ &
$L(\mathrm{d\le1'})$ &
$f_{\nu}(\mathrm{1'\le d\le3'})$ &
$L(\mathrm{1'\le d\le3'})$ &
$f_{\nu}(\mathrm{d\le d_{25}})$ &
$L(\mathrm{d\le d_{25}})$ \\
\noalign{\smallskip}
& $\mu$m & Jy &
$10^{42}\ergs$ & Jy &
$10^{42}\ergs$ & Jy &
$10^{42}\ergs$ \\
\noalign{\smallskip}
\hline
\noalign{\smallskip}
%
%
LW1 & 4.5 &
$0.6$ &
$0.4$ & 0.8 &0.6\\
%
%
LW2 & 6.7 &
$3.0$ &
$2.4$ & 2.5&2.0 & 20 & 16\\
%
%
LW3 & 14.3 &
$3.8$ &
$1.2$ & 2.9& 0.9& 21 &6.6
\\
%
%
LW4 & 6.0 &
$1.8$ &
$0.8$ &1.8&0.8\\
%
%
LW5 & 6.8 &
$2.0$ &
$0.4$ &2.4&0.4\\
%
%
LW6 & 7.7 &
$4.1$ &
$1.4$ &4.4&1.5\\
%
%
LW7 & 9.6 &
$2.4$ &
$0.8$ &2.1&0.7\\
%
%
LW8 & 11.3 &
$4.6$ &
$0.6$ & 4.2&0.6\\
%
%
LW9 & 14.9 &
$3.8$ &
$0.4$ &2.2 &0.2\\
\noalign{\smallskip}
\hline
\end{tabular}
\end{flushleft}
\end{table*}


\subsection{The ISOCAM strip maps: $\lambda$4.5, 7.7 \& 15\m}   \label{lw9overlw6sect}

In general, the strip images reveal similar morphological features as
seen in the LW2 and LW3 images. The measured fluxes and luminosities for
the individual ISOCAM filters are given in Table~\ref{lwfluxes}.

In the following, we will focus on the LW1 ($\lambda$4.0--5.0\m), the
LW6 ($\lambda$7.0-8.5\m) and the LW9 ($\lambda$14.0--16.0\m) strip maps.

\begin{figure}
\includegraphics[bb = 76 256 479 666,width=8.8cm,clip=]{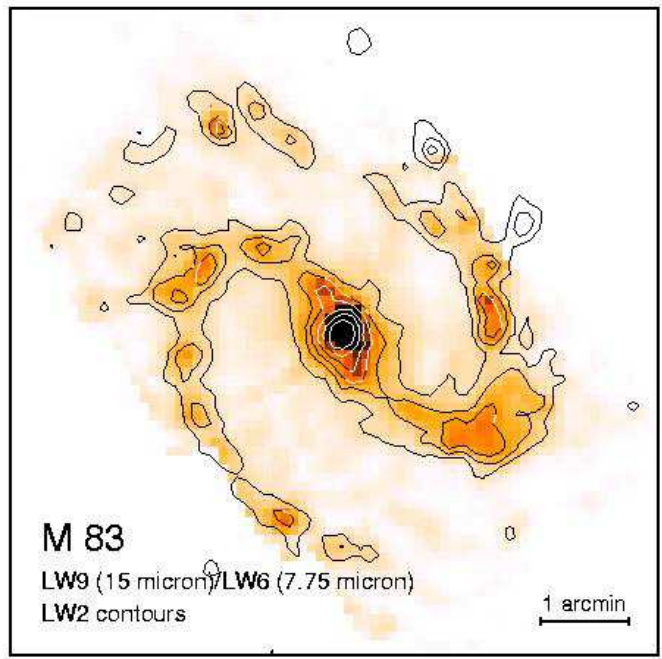}
\caption{\label{lw9-over-lw6}
Map of the ratio of LW9/LW6 ($\lambda$15.0\m\, continuum /$\lambda$7.75\m\, PAH).
}
\end{figure}

We show a LW9/LW6 ratio map in Fig.~\ref{lw9-over-lw6}. The ratio map,
overlaid with the LW2 (central wavelength: $\lambda$6.75\m) contours, shows
enhanced LW9/LW6 ratios toward regions of high star-formation activity, for
example, in the nuclear region and spiral arms. Regions of peaked LW2
emission, which traces the cluster of PAH bands show, in general,
enhanced LW9/LW6 ratios. The enhanced ratios might be interpreted in
terms of (a) increasing densities of the very small grains (VSGs)
traced by the continuum of LW9 and (b) possibly increasing
temperatures of these grains, if they are in thermal equilibrium. In
reality, the VSGs represent grains with a size distribution, and
temperature distribution (D\'esert et al.\ \cite{desert+90}). On
Galactic scales, the hot dust grain emission peaks towards the
\ion{H}{ii} regions, where the PAHs (traced by the LW6 band) show a
minimum (e.g. Cesarsky et al.\ \cite{cesarsky+96b}; Verstraete et al.\
\cite{verstraete+96}; Henning et al.\ \cite{henning+98}).  PAHs are
seen to peak in the photodissociation regions around molecular clouds
in the vicinity of star-formation regions.

\begin{figure}
\includegraphics[bb = 76 256 479 666,width=8.8cm,clip=]{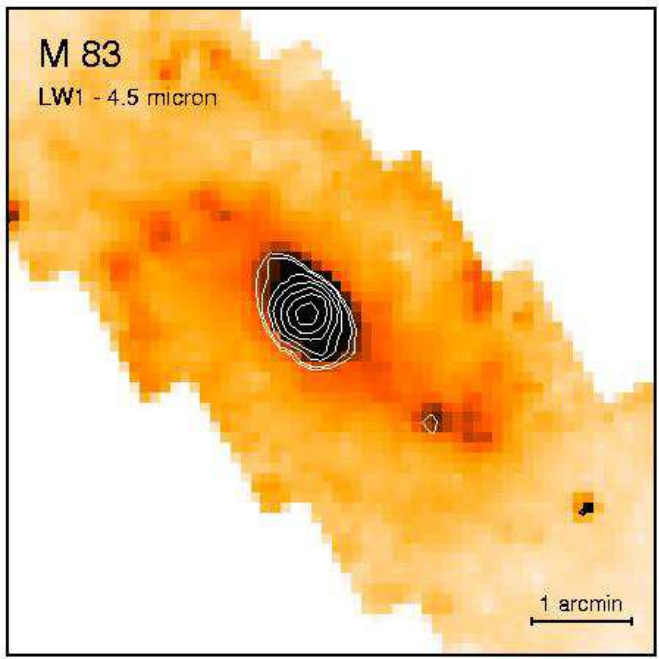}
\caption{\label{lw1}
ISOCAM LW1 ($\lambda$4.5 $\mu$m) image;
regions of higher intensity have been sketched with contours.
}
\end{figure}

We also show the LW1 strip map which traces the evolved stellar
contribution at $\lambda$4{\m}--5{\m} (Fig.~\ref{lw1}).
\label{thlab} While the bar is also obvious in the $\lambda$4.5\m\,
image, the contrast is not as sharp as in the LW2 or LW3 image,
demonstrating the evolved stellar component of the bulge being more
dominant over the narrow star-forming bar component (compare
Fig.~\ref{lw1} with Figs.~\ref{lw21w3}a and b).
The differences between the LW2 and LW1 images speak
against a strong contribution of stars ($<$ 20\% for such a late type galaxy such as Sc) to the LW2 image,
as has been argued before (e.g. Boselli et al.\ \cite{bos+03}).

\section{The radio view of M\,83}

\label{radioinf}

The $\lambda 6$~cm radio continuum map (total intensity) derived from
VLA and Effelsberg data is presented in Fig.~\ref{lw21w3}c. The radio
emission is dominated by the non-thermal synchrotron process. Neininger
et al. (\cite{neininger+93}) found that the ratio between large-scale
non-thermal and thermal emission at $\lambda 2.8$~cm varies from
$\simeq 3$ in the inner region of M\,83 to $\simeq 10$ in the outer
parts. However, the thermal
fraction may increase locally in star-forming regions of spiral arms.
Our new separation of thermal and non-thermal
emission at $23\arcsec$ resolution (see Sect.~\ref{corr})
reveals a thermal fraction at $\lambda 6$~cm which is 25\% on average,
but up to 70\% locally in spiral arms. Very similar results were obtained
for the spiral galaxy NGC~6946 (Walsh et al.\ \cite{walsh+02}).

The synchrotron intensity, $I_\mathrm{sync}$, is proportional to
$N_\mathrm{CRE} B^{1+\alpha_\mathrm{n}}$ where $N_\mathrm{CRE}$ is the
energy density of cosmic-ray electrons in the relevant energy range,
$B$ is the strength of the total field component in the sky plane, and
$\alpha_\mathrm{n}$ is the non-thermal spectral index (defined as
$I_{sync} \propto \nu^{-\alpha_\mathrm{n}}$).  In case of (often
assumed) energy density equipartition between cosmic rays and magnetic
fields, $I_\mathrm{sync} \propto B^{3+\alpha_\mathrm{n}}$. Thus
synchrotron intensity is a sensitive measure of field strength and its
variations.

Emission from the bar is concentrated toward the outer edge
(i.e. leading with respect to the galaxy's rotation). This can be
regarded as a signature of a shear shock due to gas infall on the
inner side of the bar (see numerical models e.g. by Athanassoula\
\cite{athana92}).  If the magnetic field is compressed together with
the gas, a compression ratio of 4 in a strong adiabatic shock should
result in an increase in synchrotron intensity of a factor of 16 if
the density of cosmic-ray electrons is constant (assuming
$\alpha_\mathrm{n} \simeq 1$, see Sect.~\ref{corr}). This factor could
be larger if the cosmic-ray density also increases in the shock
(energy equipartition).  However, the intensity contrast between the
radio bar and its surroundings is only $\simeq 3$, similar to
NGC~1097, a galaxy with a larger bar (Beck et al.\
\cite{beck+99}). Either the shock generates magnetic turbulence, or
the magnetic field is not frozen into the gas flow, or the shock has a
complex 3-D structure. Better numerical models, which take magnetic
fields into account, are required.

\begin{figure}
\includegraphics[bb = 76 256 479 666,width=8.8cm,clip=]{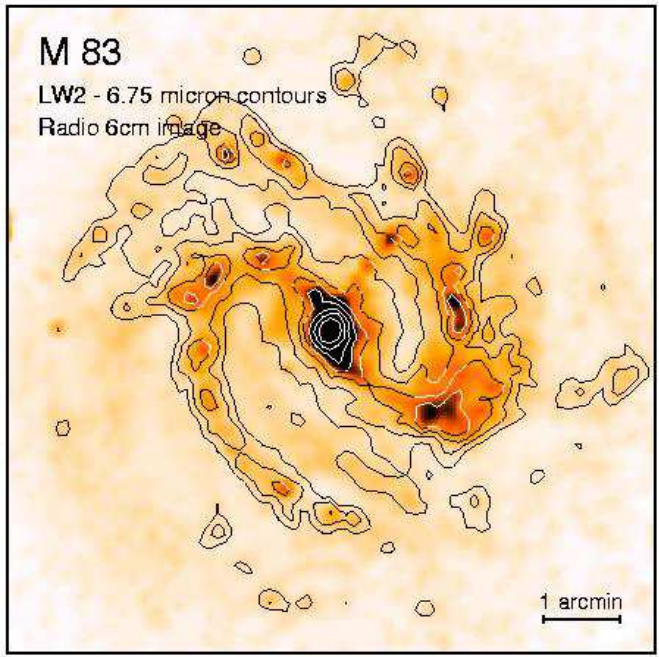}
\caption{ \label{lw2-over-radio}
Contours of the ISOCAM LW2 ($\lambda$6.75{\m} central wavelength; angular
resolution $6\arcsec$) image overlaid onto the $\lambda$6~cm radio continuum
image (resolution $10\arcsec$).
}
\end{figure}

Fig.~\ref{lw2-over-radio} demonstrates the close association between
the radio emission and the MIR dust emission.  This is most obvious
in the spiral arm 2 to 3$\arcmin$ north of the center, where both the
radio continuum and dust distribution reveal a sharp edge while the
optical spiral arm extends much further to the north.

Three sources, which are not obvious in the MIR images but appear in the
radio map at $0\farcm4$--$1\farcm4$ north-west of the center
(Fig.~\ref{lw2-over-radio}), are supernova remnants and/or background
quasars. Their positions do not coincide with any of the observed
20th century supernova events.
In principle, supernova remnants and dusty quasars have been detected at MIR
wavelengths with ISOCAM (e.g. Lagage et al.\ \cite{lagage+96}; Arendt et
al.\ \cite{arendt+99}; Leech et al.\ \cite{leech+01}).  While it is
conceivable that a hint of MIR emission is detected toward these radio
sources, they are not detectable at the sensitivity and angular
resolution of these images.

The close correlation between radio continuum and mid-infrared dust
emission is indeed striking in M\,83 (Fig.~\ref{lw2-over-radio} and
Figs.~\ref{lw21w3}a, b, c) and in NGC~6946 (Frick et al.\
\cite{frick+01}; Walsh et al.\ \cite{walsh+02}).
The explanation is far from being trivial (see Sect.~\ref{mirradio}).

\section{Comparison of the MIR morphology with further tracers of
the interstellar medium}

\subsection{H$\alpha$ emission}
\label{halphamorph}

\begin{figure}
\includegraphics[bb = 76 112 451 480,angle=-90,width=8.8cm,clip=]{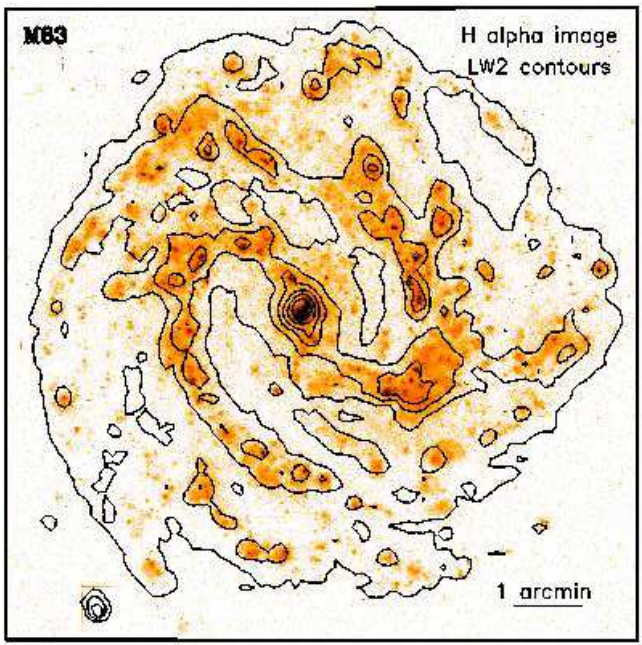}
\caption{ \label{lw2-over-halpha}
LW2 ($\lambda$6.75\m\ central wavelength; angular resolution $6\arcsec$) contours
overlaid onto the H$\alpha$ image of Lundgren et al.(\cite{lundgren+05}).
}
\end{figure}

The H$\alpha$ image of M\,83 was initially presented in Lundgren et al. (\cite{lundgren+05}). The FWHM of the PSF is $1\farcs1 8$ (30~pc). The calibrated
H$\alpha$ image (Fig.~\ref{lw21w3}d) is used here for quantitative
comparison with the radio and MIR emission, both of which have spatial resolutions {poorer} than that of H$\alpha$.

Fig.~\ref{lw2-over-halpha} presents the LW2 contours
overlaid on the H$\alpha$ image of Lundgen et al. (\cite{lundgren+05}).
In general, the MIR emission and H$\alpha$ emission highlight the
same region, but differences exist, e.g. in the north-east
(Sect.~\ref{radioinf}).
The spiral arms are traced well at both wavelengths, and many
of the bright H$\alpha$ emission regions correspond to local maxima of
the MIR emission. Diffuse H$\alpha$ emission is also present throughout the interarm region. The correlation between MIR and H$\alpha$
is discussed in Sec.~\ref{miralpha}.

\subsection{Diffuse X-ray emission}

\begin{figure}
\includegraphics[bb = 26 27 443 446,width=8.8cm,clip=]{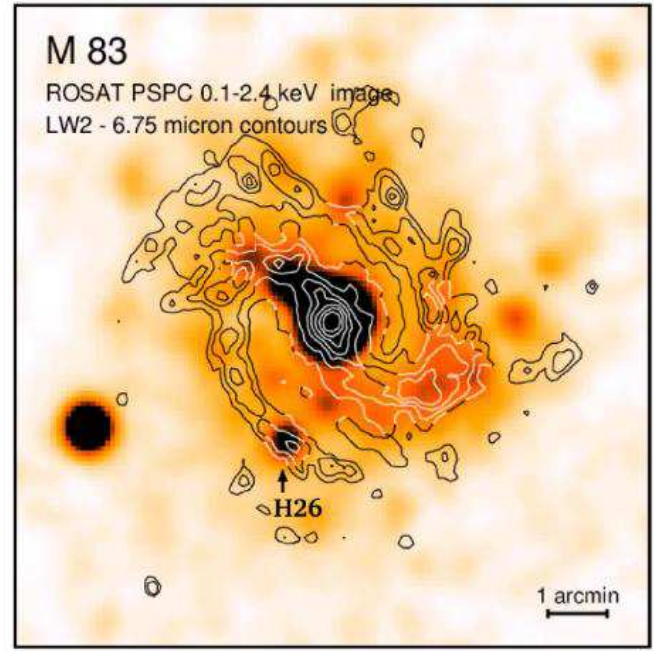}
\caption{ \label{lw2-over-rosat}
LW2 ($\lambda$6.75\m\ central wavelength; angular resolution
$6\arcsec$) contours overlaid onto the ROSAT PSPC 0.1--2.4~keV image
from Ehle et al.  (\cite{ehle+98}) with an energy-dependent angular
resolution of 25$\arcsec$--52$\arcsec$.  H26, labeled in the figure, is
an X-ray luminous source coinciding with a MIR peak in a spiral arm.
It is identified as a supernova remnant candidate (Ehle et al.\
\cite{ehle+98}).}

\end{figure}

M\,83 was observed with the {\it Einstein} (Trinchieri et al.\
\cite{trinchieri+85}), GINGA (Ohashi et al.\ \cite{ohashi+90}),
ASCA (Okada et al.\ \cite{okada+97}) ROSAT (Ehle et al.\
\cite{ehle+98}; see Fig.~\ref{lw2-over-rosat}; Immler et al.\
\cite{immler+99}) and {\it Chandra} (Soria \& Wu\ \cite{soria+wu02};
Soria \& Wu\ \cite{soria+wu03}) observatories. Diffuse X-ray emission
was detected by ROSAT as well as by {\it Chandra}. From spectral
investigations, roughly 65\% of the diffuse X-ray emission was
attributed to a million degree component of the ISM, while the
remaining part was attributed to unresolved point sources.
While {\it Chandra} was well suited to clearly resolve the
starburst nucleus region (see Sect.\ref{blobnucleus}) for the first time and to detect the discrete
source population in M\,83, the diffuse extended emission could be traced (after
spatial smoothing) only from the optically bright spiral arms.

ROSAT PSPC, in contrast, showed extended X-ray radiation from almost the whole
optically visible part of that galaxy. Ehle et al. (\cite{ehle+98}) interpreted
the diffuse X-ray emission as a multi-temperature thermal plasma: a soft X-ray
emitting spherical halo component and a hotter disk component, dominating the
emission in the harder 0.5--2.0 keV ROSAT band.

\subsection{Molecular gas}

\begin{figure}
\includegraphics[bb = 76 121 451 480,angle=-90,width=8.8cm,clip=]{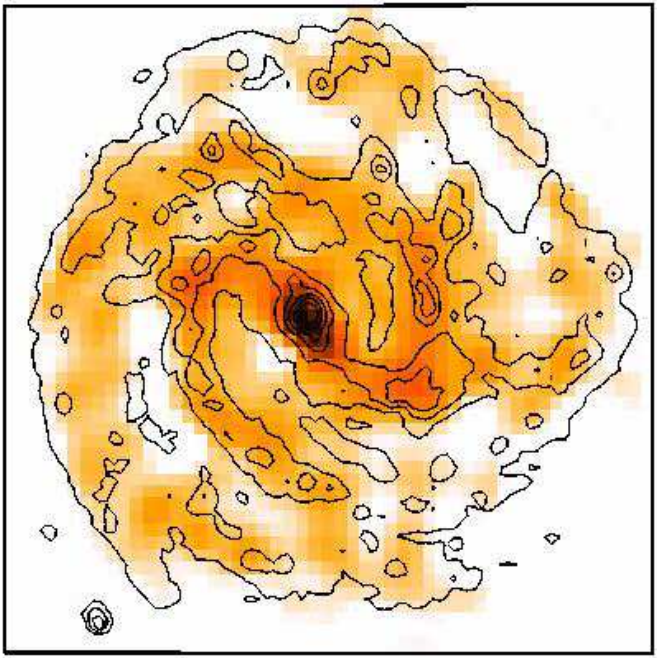}
\caption{ \label{lw2-over-CO10}
LW2 ($\lambda$6.75\m\ central wavelength; angular resolution $6\arcsec$) contours 
onto the \mbox{CO($J$=1--0)} image of Lundgren et al. (\cite{lundgren+04a}) with a resolution of $22\arcsec$.
}
\end{figure}

Comparison of the distribution of the \mbox{CO($J$=1--0)} emission at
22\arcsec\ angular resolution (Lundgren et al.\ \cite{lundgren+04a})
with that of the MIR dust emission shows very good spatial agreement
in the disk (see Fig.~\ref{lw2-over-CO10}).  However, there are also
displacements: local peaks (or in some cases crests) of
\mbox{CO($J$=1--0)} emission appear to be shifted about 10--20\arcsec,
but there is no apparent systematic trend such that the \mbox{CO($J$=1--0)} emission lies
outside the MIR emission or vice verse.  The local maxima of the MIR
emission are better aligned with those of the \mbox{CO($J$=2--1)}
emission with 14\arcsec\ resolution (Lundgren et al.\ \cite{lundgren+04a}).
This is perhaps not so surprising since
molecular gas clouds in regions with active star formation are likely
to have an increased kinetic temperature (and perhaps also higher
density), resulting in stronger emission from higher-$J$ emission
lines.

The central peak in the MIR data agrees well with the IR center position
(Gallais et al.\ \cite{gallais+91}). The nucleus appears as a single peak in the MIR
data and does not have the double-peaked nature seen in the \mbox{CO($J$=1--0)}
(Handa et al.\ \cite{handa+94}),
\mbox{CO($J$=2--1)} (Lundgren et al.\ \cite{lundgren+04a}) and CO($J$=3--2 and $J$=4--3)
data (Petitpas \& Wilson\ \cite{petit+wilson98}). See Sect 6.2 for further discussion of the nuclear region.

Additionally, the nuclear and bar region, as well as the south-eastern spiral arm,
were observed in CO line emission with high resolution (Handa et al.\ \cite{handa+90};
Lord \& Kenney\ \cite{lord+kenney91}; Rand et al.\ \cite{rand+99};
Dumke et al.\ \cite{dumke+05}).
These observations will be discussed in Sections \ref{blobnucleus}
and \ref{sespiralarm}, respectively.

\section{Discussion}

\subsection{Radial brightness profiles of different tracers of the
interstellar medium}
\label{radialprofiles}

\begin{figure*}
\includegraphics[bb = 42 33 579 722,angle=-90,width=13cm]{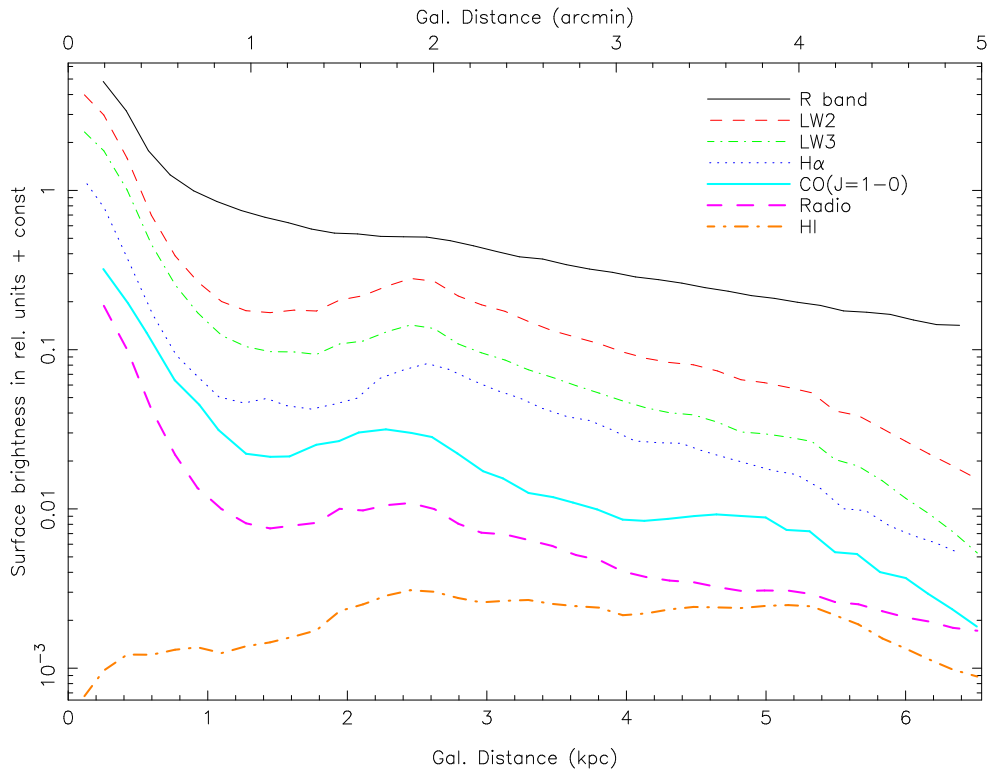}
\caption{ Radial intensity profiles versus galactocentric
radius for different tracers in M\,83.}
\label{radial-profiles}
\end{figure*}

 The radial extent of different tracers of the neutral, molecular and
ionized ISM was compared. Figure~\ref{radial-profiles} shows the
profiles for seven tracers:
R band (Larsen \& Richtler\ \cite{larsen+99}), MIR (LW2 and LW3), H$\alpha$ (Lundgren et al.\ \cite{lundgren+05}),
\mbox{CO($J$=1--0)} (Lundgren et al.\ \cite{lundgren+04a}), radio continuum ($\lambda$6~cm), and \ion{H}{i} (Tilanus \& Allen\ \cite{tilanus+allen93}).
The profiles have been calculated from $23\arcsec$ resolution maps, which were
corrected for distortion due to the inclination of the
galaxy (24$^{\circ}$), by adjusting the distance scale along the minor axis by 1/cos(24).

For all profiles, except that of \ion{H}{i}, the intensity declines from the
central region out to $r\simeq1\arcmin$, where most profiles have a
local minimum and a corresponding local maximum at $r\simeq2\arcmin$.
 In the radial range $2\arcmin$--$4\arcmin$ all tracers, except \ion{H}{i}, show an
exponential decline. The MIR and H$\alpha$ profiles seem to be
similar, except that the MIR profiles near $2\arcmin$ have more pronounced peaks, and beyond $4\arcmin$ (5 kpc), the MIR profile shows a steeper decline. We find a scale length of $1\farcm3\pm0\farcm1$
($\cor1.7\pm0.1$~kpc) for the H$\alpha$ and the MIR.  The same value was found for the thermal radio
continuum emission at $\lambda$2.8~cm (Ehle et al.\ \cite{ehle+98}).
These authors also found the same scale length ($1\farcm3\pm0\farcm1$)
for the hard X-ray emission from the disk for $r>2\farcm1$ ($r>3$~kpc).

In the optical wavelength region (here represented by the R band),
\mbox{CO($J$=1--0)} and total radio continuum, the scale lengths are
considerably larger: $2\farcm2\pm0\farcm1$, $2\farcm1 \pm 0\farcm3$\
and $1\farcm7\pm0\farcm1$, respectively.
The fact that H$\alpha$ has a smaller scale length than that of the stellar light
has previously been noted in
M\,83 by Larsen \& Richtler (\cite{larsen+99}). According to Ryder \& Dopita\ (\cite{ryder+94}), however, the H$\alpha$ scale length in galaxies is usually larger that that of
the stellar broad-band emission. The explanation for M\,83 probably
lies in the fact that the central region and the inner spiral arms are
undergoing bursts of star formation.


For $r>4\arcmin$ ($r>5$~kpc) the decline of the MIR and H$\alpha$
emission steepens while that of the R band and radio continuum continues
without a break.
The scale lengths in the radial region $4\farcm0$--$5\farcm0$
are $0\farcm7\pm0\farcm1$, $0\farcm6\pm0\farcm1$ and $0\farcm9\pm0\farcm1$
for LW2, LW3 and H$\alpha$, respectively, while the R band and radio
continuum have scale lengths of $2\farcm6\pm0\farcm2$ and $1\farcm7\pm0\farcm1$,
similar to those in the inner disk.
The scale length of the \mbox{CO($J$=1--0)} cannot be determined reliably
since this radial range lies partially outside the area which was covered
in Lundgren et al.\ (\cite{lundgren+04a}).
The large scale length in radio continuum is known to exist in
many spiral galaxies, e.g. in NGC~6946 (Walsh et al.\ \cite{walsh+02}),
and is an effect of the diffusion of cosmic-ray electrons away from the
star-forming regions and the large extent of magnetic fields.

For \ion{H}{i}, the decline begins only beyond 5.4 kpc and is in good agreement with the existence of a giant
($r>40³\arcmin$) \ion{H}{i} halo for M\,83 (Huchtmeier \& Bohnenstengel\
\cite{hucht+bohnen81}).

In the nuclear region all of the tracers except \ion{H}{i} reach
brightness maxima which are most pronounced in H$\alpha$ and the
radio. Active star formation can convert \ion{H}{i} into
molecular and ionized gas. From a comparison of the MIR and radio
data, we conclude that we have a higher radio/MIR intensity ratio for
the nucleus than for the disk. While a part of this effect may be due
to ISOCAM saturation effects, we may also argue for a compression of
the magnetic field by the bar shock, enhancing the radio (synchrotron)
emission. Alternatively, the coupling between magnetic fields and gas
clouds may be more effective in the nuclear region, e.g. due to higher
gas temperatures. Such a higher temperature is indicated by increasing
CO($J$=4--3)/CO($J$=3--2) line ratios in the nuclear region (Petitpas
\& Wilson\ \cite{petit+wilson98}).

The steeper decline of the MIR intensity versus the optical or
\ion{H}{i} brightness in our data is in good agreement with results of
Roussel et al. (\cite{roussel+01a}) who find the MIR disks in spiral
galaxies are smaller than the optical or \ion{H}{i} disks. We further
confirmed this trend with the help of the MIR strip observations which
cover $17\farcm4 \times 5\farcm4$, versus the $12\farcm 8\times
12\farcm 8$ field of the LW2/LW3 observations. Therefore, our findings
cannot be explained in terms of difficulties of the background estimation
in case of the LW2 and LW3 maps which almost completely cover the optical
extent of M\,83 ($d_{25}=11\farcm5$).

\subsection{The nuclear and bar region}
\label{blobnucleus}

To further understand the nature of the nuclear MIR and radio
emission, we zoom into the central 2'x2' region ($2.6\kpc \times
2.6~\kpc$) of M\,83 and investigate their relationship with various ISM
tracers, such as H$\alpha$, \ion{H}{i},
\mbox{CO($J$=1--0)}, \mbox{CO($J$=2--1)} and \mbox{CO($J$=3--2)}
(Fig.~\ref{nucreg}). We have also {sketched in
Fig.~\ref{nucreg}f the NIR emission features (Adamson et al.\
\cite{adamson+87}; Gallais et al.\ \cite{gallais+91} Sofue \& Wakamatsu\ \cite{sofue+94}; Elmegreen et al.\
\cite{elmegreen+98}; Thatte et al.\ \cite{thatte+00}), namely the starburst
shells, the double nucleus and the mini-bar.

\begin{figure*}
\centerline{\includegraphics[bb = 103 127 493 702,width=14cm,clip=]{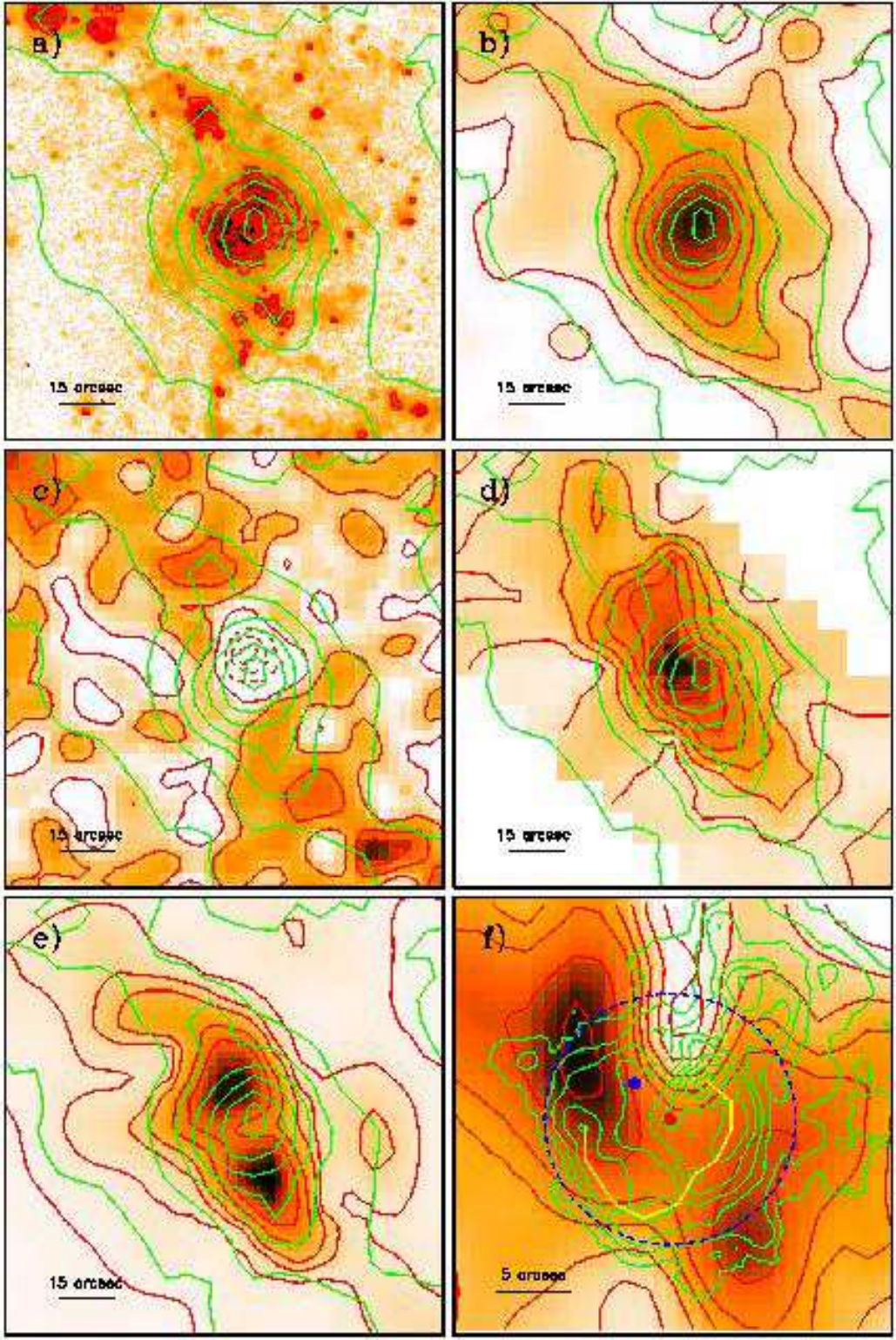}}
\caption{\label{nucreg}
The nuclear region of of M\,83.
The scales of the figures a) to e) are identical, while figure f) is zoomed
in by a factor of about 3 with respect to the other images.
In figures a) to e) the green contours show the ISOCAM LW2 data
($\lambda$6.75{\m} central wavelength, resolution $6\arcsec$), while
the background image and the red contours represent the following:
\ \ {\it (a)} H$\alpha$ emission (logarithmic color scale, angular
resolution $1\farcs1$, Lundgren et al.\ \cite{lundgren+05})
\ \ {\it (b)} $\lambda$6~cm radio continuum emission (logarithmic color scale,
resolution $10\arcsec$)
\ \ {\it (c)} \ion{H}{i} column density (Tilanus \& Allen\
\cite{tilanus+allen93}, linear color scale, resolution $12\arcsec$).
Note that the \ion{H}{i} line is seen in absorption
towards the center of the galaxy. This is represented with dashed contours.
\ \ {\it (d)} \mbox{CO($J$=1--0)} emission
(Handa et al.\ \cite{handa+90}, linear color scale, resolution $16\arcsec$).
\ \ {\it (e)} \mbox{CO($J$=2--1)} map (Lundgren	 et al.\ \cite{lundgren+04a},
linear color scale, resolution $14\arcsec$).
\ \  In image {\it (f)} the green contours represent H$\alpha$ emission,
and they are superposed onto an CO(3-2) map
(Petitpas \& Wilson\ \cite{petit+wilson98}, linear color scale,
resolution, resolution $14\arcsec$).
The yellow line sketches out the star forming arc.
The blue and the red dots are the bright peak seen in the K band,
and the center of symmetry of the K band isophotes,
respectively (Thatte et al.\ \cite{thatte+00}).
The blue dashed circle represents the ring seen in the J-K images
(Elmegreen et al.\ \cite{elmegreen+98}).}
\label{fig:nuclear_zoom}
\end{figure*}

From the H$\alpha$ morphology (Fig.~\ref{nucreg}a), it is
clear that the bulge region is not a point source, but shows two
``bubble-like'' structures opening out to the north-west and south-east of the nucleus.
While the MIR image cannot resolve the nuclear emission region due to
its lower angular resolution, there is good
coincidence of MIR and H$\alpha$ emission on a larger scale toward the central region. Both
tracers peak at the nucleus of the galaxy.  Bright H$\alpha$ maxima
at the position angle of $10\degr$ and $190\degr$ at distances from the nucleus
of $\sim 30\arcsec$ (650~pc) and $\sim 20\arcsec-30\arcsec$,
respectively, show that the optical bar alignment and star formation
activity along and at the ends of the bar are traced by ``tongues'' of the LW2 contours.  The LW2 contours also outline the compact H$\alpha$ regions at the position angle of $270\degr$ and
$335\degr$ (distances of $\sim 40\arcsec$ and $\sim 35\arcsec$,
respectively).

Fig.~\ref{nucreg}b displays the $\lambda$6~cm radio continuum image and red
contours and the LW2 contours in green. The MIR observations and the radio observations do not provide
sufficient angular resolution to detect a possible correlation with the
NIR emission components (sketched in Fig.~\ref{nucreg}f).
 The nucleus itself is not resolved in either image, but``tongue-like''
structures of the contours from the nucleus following the bar to the
south and north are visible in both bands.

A correspondence of the \ion{H}{i} void in the nuclear area (dashed red contours) and the
MIR emission peak is shown in Fig.~\ref{nucreg}c.
The tongue-like LW2 emission features (green contours) south-west and north-east of the
nucleus coincide with local \ion{H}{i} maxima.

Our Fig.~\ref{nucreg}d
shows the \mbox{CO($J$=1--0)} image and red contours (Handa et al.\ \cite{handa+90})
in comparison with LW2 contours (green).
Both tracers peak on the nucleus
of the galaxy. The \mbox{CO($J$=1--0)} contours do not show the ``bottleneck''
structure at $\sim 30\arcsec$ (650~pc) NE and SW of the nucleus. If
real, and not an effect due to the large spacing ($15\arcsec$) of the CO
map, this might further indicate that we indeed trace different components:
the molecular gas along the bar and the MIR PAH emission along the optical
spiral arms and leading edge of the bar.

As shown in our overlay of the MIR contours (resolution $6\arcsec$) on the
CO($J$=2--1, resolution $14\arcsec$) data (Fig.~\ref{nucreg}e), we can
exclude the presence of a double-peaked MIR structure separated by
$\sim 20\arcsec$ (430~pc) as detected in the
\mbox{CO($J$=2--1)} (Lundgren et al.\ \cite{lundgren+04a}) and
\mbox{CO($J$=3--2)} (Petitpas \& Wilson\ \cite{petit+wilson98}) data. The MIR peak appears to lie in the middle of the CO peaks.
Similar to the \mbox{CO($J$=2--1)}, the contours of the
LW2 emission are located at the leading edge of the bar, which argues
for the presence of dust lanes along the leading edges of the
bar. Recent CO($J$=3--2) data from the  Heinrich Hertz Telescope (HHT) (Dumke et al.\ \cite{dumke+05}) also show this.  These results confirm the conclusion of Elmegreen et al.\ (\cite{elmegreen+98}) who have demonstrated this point using J-K color images.

To obtain further insight on the nature of the nuclear emission beyond
the resolution of the MIR and radio continuum data, we overlaid the
H$\alpha$ contours on the CO($J$=3--2) image (Fig.~\ref{nucreg}f), zooming in further.
 In this figure we sketched the two near-infrared nuclei (blue and red dots), the star forming arc (yellow) as well as the near infrared ring (blue-dashed ring) seen in the J-K images (Elmegreen et al.\ \cite{elmegreen+98}). The
brightest H$\alpha$ emission peak is close to the
bright near-infrared peak of the galaxy (blue dot). Both H$\alpha$ emission peaks lie on the ring seen in the J-K images, while the center of symmetry of the K-band emission lies at the center of the image (red dot; Thatte et al.\ \cite{thatte+00}), between the two H$\alpha$ emission peaks.
Optical spectroscopy of these metal-rich \ion{H}{ii}
regions is given in Bresolin \& Kennicutt (\cite{bresolin+kennicutt02}).
For the emission peak at the position of the  north-eastern nucleus, a
Wolf-Rayet bump is detected and the age of the \ion{H}{ii} region
is approximately 5~Myr.

Two bubble-like structures of ionized gas are visible to the NW and SE,
both oriented nearly perpendicular to the major axis and bar of M\,83.
These structures are embedded
in large-scale, diffuse H$\alpha$ emission with an extent of
$18\arcsec\times 12\arcsec$ ($390\pc \times 260\pc$).
While star-forming regions of different intensities or a double or
multiple nucleus can, in principle, explain the morphology of the
H$\alpha$ emission peaks, one might be tempted to speculate about the
detection of an inflow- or outflow/superwind-like H$\alpha$ morphology
in the central region of the galaxy as known to exist for other galaxies with
nuclear starburst activities (e.g. Bland \& Tully\
\cite{bland+tully88}; Lehnert et al.\ \cite{lehnert+99} for M~82, or
Schulz \& Wegner\ \cite{schulz+wegner92} for NGC~253).  Dense gas and dust present in spiral arms, compressed by spiral density waves can also affect the morphology of the H$\alpha$ emission.

The nuclear region of M\,83 could clearly be resolved with the advent
of {\it Chandra} observations (Soria \& Wu\ \cite{soria+wu02}). A highly structured
composition of 15 discrete sources and unresolved clump-like features due to
faint sources or hot gas clouds was discovered. The nuclear X-ray emission is
strongest inside the outer dust ring (Elmegreen et al.\ \cite{elmegreen+98})
and contains both the IR photometric nucleus (consistent with the second
brightest {\it Chandra}-detected nuclear X-ray source) and the star-forming arc (in
the south). While the optical emission is more strongly concentrated around
the IR nucleus and along the star-forming arc, the X-ray emission is
more uniformly distributed and extends to larger galactocentric radii: to the
south-west along the main galactic bar and also towards the north-west across
a dust lane (part of the outer circumnuclear ring) and in the direction
perpendicular to the bar. X-ray emission, uncorrelated with any bright
(star-forming) optical regions, was also found to the east of the IR photometric
nucleus and might originate from remnants of stellar evolution. The X-ray
spectrum of the unresolved nuclear X-ray emission suggests that the ISM has been
 enriched by the ejecta of Type II supernovae and stellar winds (see also Soria \& Wu\ \cite{soria+wu02}).

\subsection{The south-eastern spiral arm}
\label{sespiralarm}

The south-eastern spiral arm of M\,83 was studied in detail at several
wavelengths. The atomic and ionized hydrogen in this arm region was
observed by Tilanus \& Allen (\cite{tilanus+allen93}). Deutsch \& Allen
(\cite{deutsch+allen93}) focused on the non-thermal radio emission and
H$\beta$ emission. \mbox{CO($J$=1--0)} observations were carried out by Lord \&
Kenney (\cite{lord+kenney91}) and Rand et al. (\cite{rand+99}). Rand et
al. (\cite{rand+99}) stress, as indicated already by the observations of
Lord \& Kenney (\cite{lord+kenney91}), that the CO emission of the arm
was found to be aligned with a dust lane delineating the inner edge of
the spiral arm, but beyond a certain point along the
arm the emission shifts downstream from the dust lane. While the CO
emission at least partially covers the dust lane, the \ion{H}{i}
emission along the arm is offset from the dust lane to the east
by up to 700~pc. Tilanus \& Allen (\cite{tilanus+allen93}) propose a
diffuse shock or dissociation scenario to explain the \ion{H}{i}
findings. Rand et al. (\cite{rand+99}) discuss heating of the CO gas by
UV radiation from young stars, heating by low energy cosmic rays or an
interstellar medium of different density components,in order to explain the
offset of the CO emission from the dust lane.

\begin{figure*}
\unitlength1.0cm
\begin{picture}(9,6.63)
\thicklines
\put(0,0){\framebox(4.5,6.63){
\includegraphics[bb = 125 265 380 637,width=4.5cm,clip=]{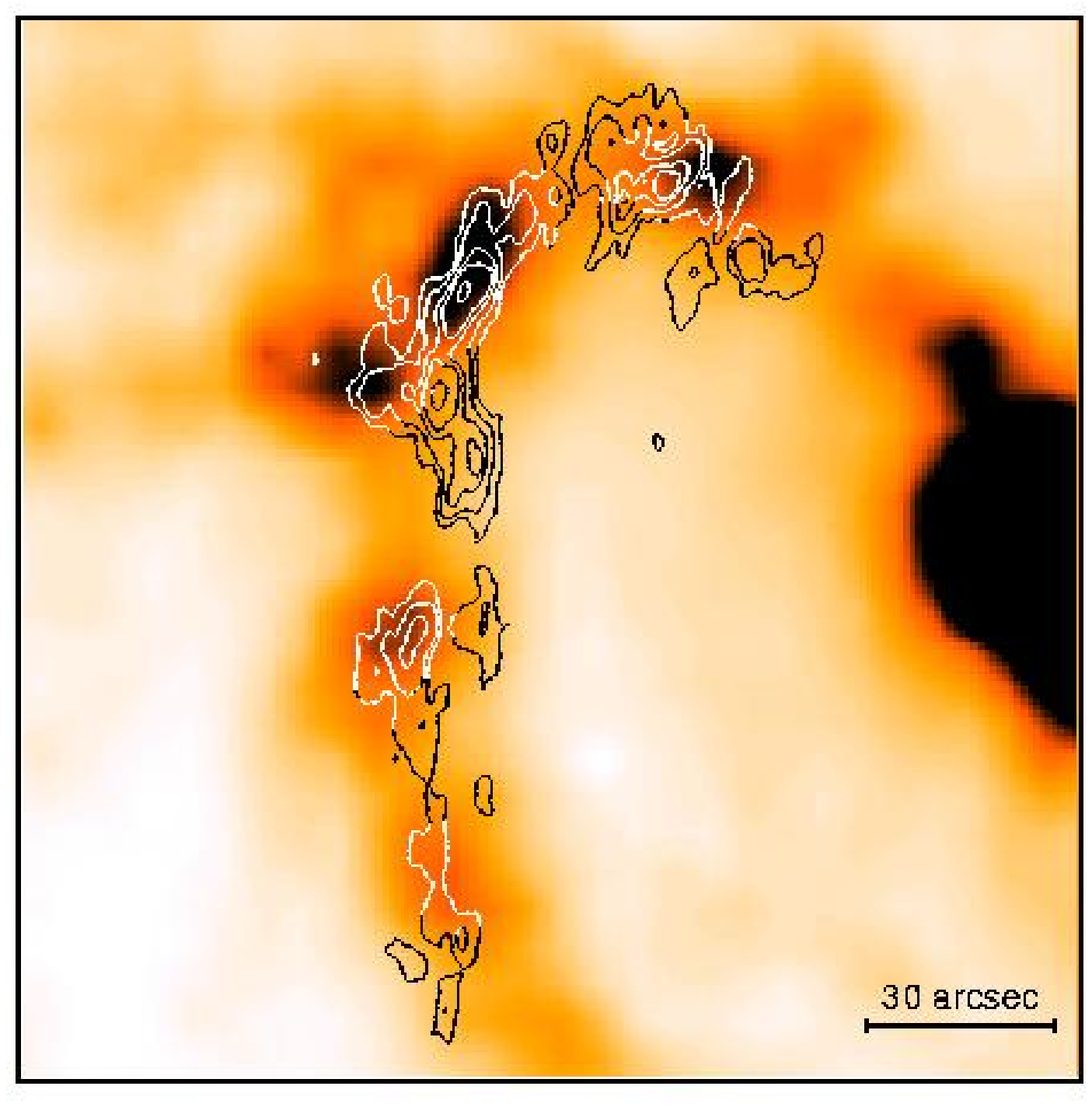}
         }}
\put(3.15,1.0){$30\arcsec$}
\put(2.5,.48){LW2 image}
\put(2.5,.15){CO contours}
\put(2.8,.9){\line(1,0){1.15}}
\put(4.5,0){\framebox(4.5,6.63){
\includegraphics[bb = 125 265 380 637,width=4.5cm,clip=]{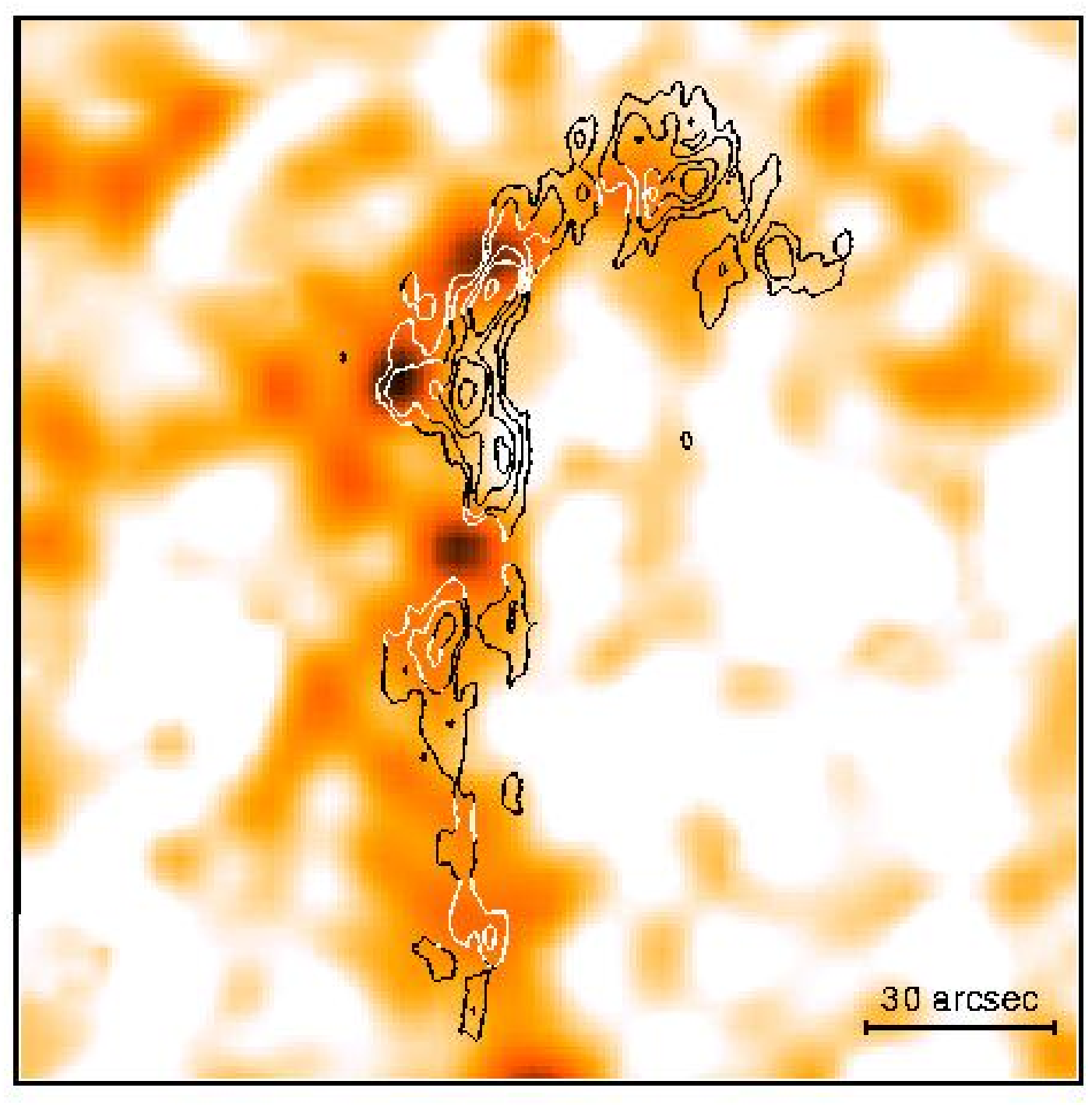}
         }}
\put(7.65,1.0){$30\arcsec$}
\put(7,.48){\ion{H}{i} image}
\put(7,.15){CO contours}
\put(7.3,.9){\line(1,0){1.15}}
\put(7.3,3.45){\vector(-1,0){.4}}
\put(9,0){\framebox(4.5,6.63){
\includegraphics[bb = 125 265 380 637,width=4.5cm,clip=]{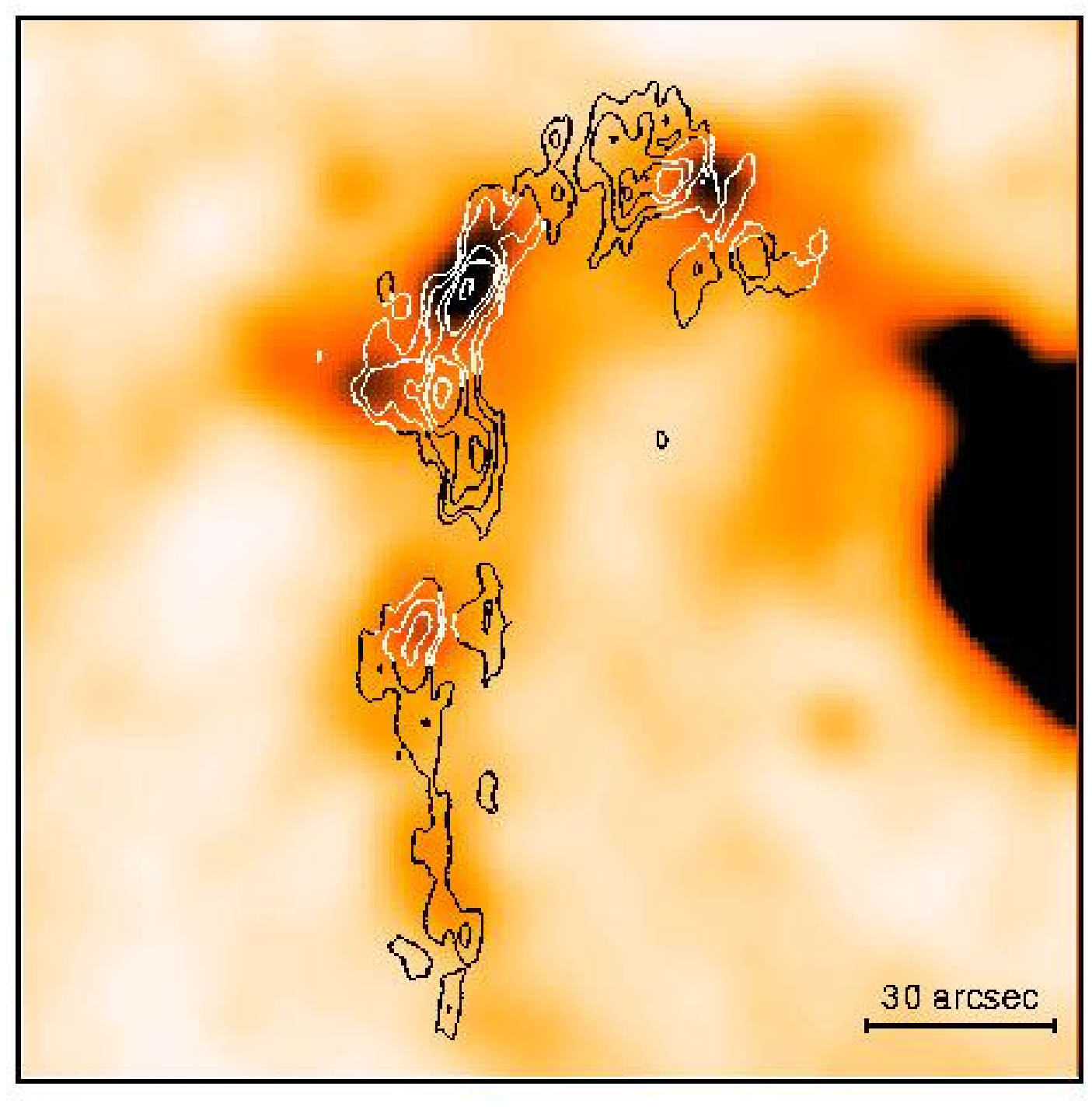}
         }}
\put(12.15,1.0){$30\arcsec$}
\put(11.5,.48){radio image}
\put(11.5,.15){CO contours}
\put(11.8,.9){\line(1,0){1.15}}
\put(13.5,0){\framebox(4.5,6.63){
\includegraphics[bb = 125 265 380 637,width=4.5cm,clip=]{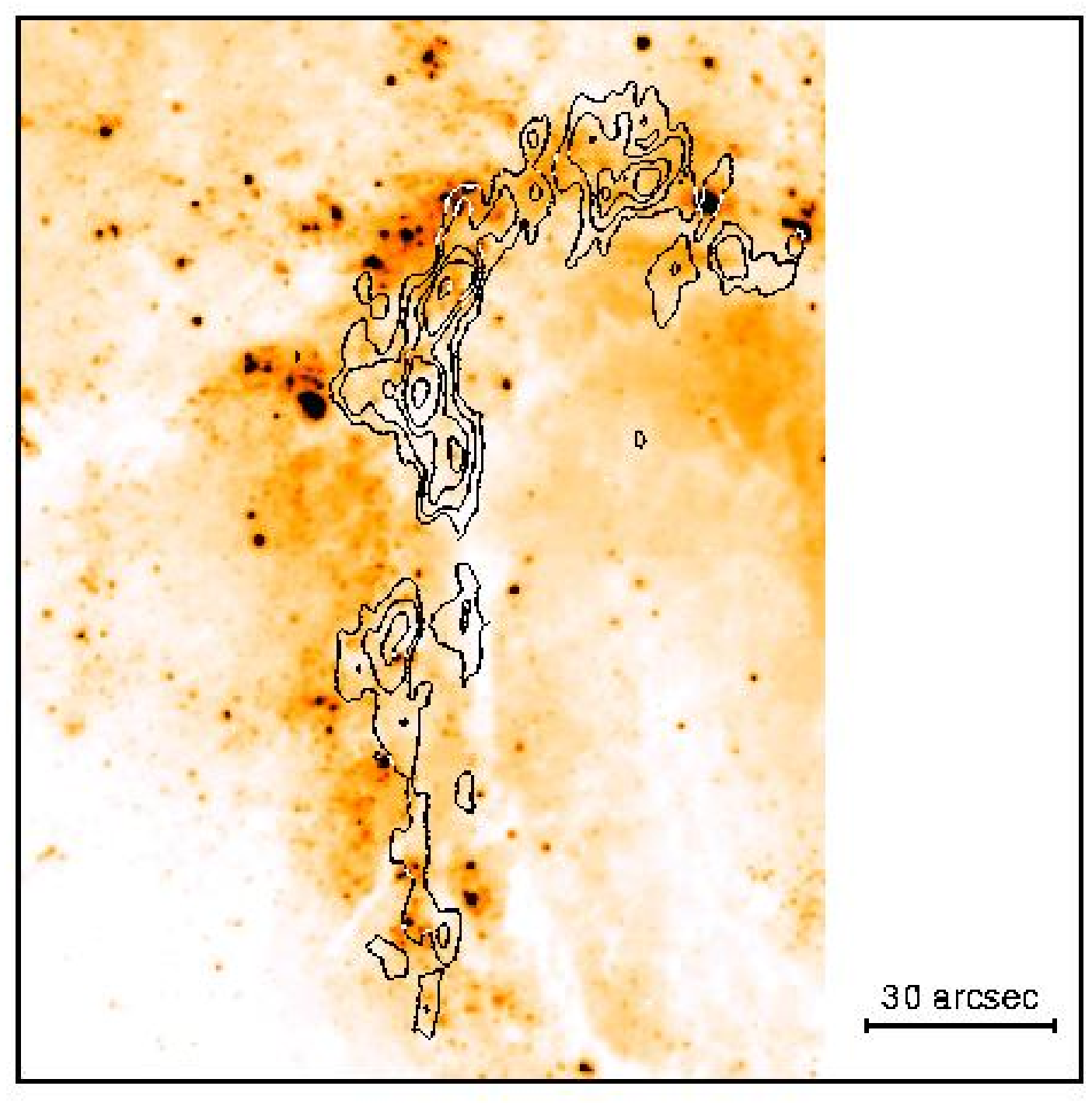}
        }}
\put(16.65,1.0){$30\arcsec$}
\put(16,.48){VLT B image}
\put(16,.15){CO contours}
\put(16.3,.9){\line(1,0){1.15}}
\put(.3,.3){\large a)}
\put(4.8,.3){\large b)}
\put(9.3,.3){\large c)}
\put(13.8,.3){\large d)}
\end{picture}
\caption{\label{co}
South-eastern spiral arm of M\,83:
{\it (a)} Contours of the CO emission  on the ISOCAM LW2 image,
{\it (b)} \ion{H}{i} image of Tilanus \& Allen (\cite{tilanus+allen93}) and contours of CO,
{\it (c)} the radio continuum map with CO contours and
{\it (d)} the VLT/FORS1 B image (ESO archive) and CO contours.
The first three images were rebinned to a pixel size of $1\arcsec$ for
this representation. The CO map is from Rand et al. (\cite{rand+99}),
and we show the same contour levels as in their publication (their
Fig.~2).
}
\label{fig:spiralarm_zoom}
\end{figure*}

In our Figs.~\ref{co}a--d, we present the contours of the CO map as
shown by Rand et al. (\cite{rand+99}) in their Fig.~2 overlaid on the
LW2, \ion{H}{i}, radio continuum and VLT/FORS1 B band (ESO archive)
images. The LW2, \ion{H}{i} and radio data were rebinned to a pixel
size of $1\arcsec$ for the comparison with the \mbox{CO($J$=1--0)} contours.
The CO emission lies along a sharp dust lane visible in blue images
of M\,83 in most of the region displayed, but near a certain point the
emission falls below the detection limit, and this point is marked by
an arrow in Fig.~\ref{co}b. It is worth mentioning that the
non-detection of CO emission coincides with a local maximum in the
\ion{H}{i} emission, corresponding to a column density of $2.2\times
10^{21}\cmtwo$ at the resolution of the \ion{H}{i} map or an
average density above $1.7\times 10^{21}\cmtwo$ for an integration
radius of $7\farcs 5$. We detect neither an enhancement in MIR
nor in radio continuum emission at this location. Further to the south,
the CO emission bifurcates, with a local maximum to the west. The
emission is aligned with a dust lane and an elongated emission
feature following the arm structure as highlighted by the highest
MIR and radio intensities. This alignment might argue for enhanced star
formation and cosmic ray densities (Rand et al.\ \cite{rand+99})
or strong magnetic fields along the CO filament extending from the
local \ion{H}{i} maximum marked in Fig.~\ref{co}b to the south.

However, the OVRO \mbox{CO($J$=1--0)} data is not a reliable source to trace the
bulk of molecular gas since it only reveals 2--5\% of the total
\mbox{CO($J$=1--0)} emission in this region (see Rand et al.\ \cite{rand+99}).
It is likely that the emission that was detected has its origin
in clouds that are heated or compressed by star formation and/or shocks,
instead of tracing the cool molecular gas.
As an example, the OVRO data traces the leading edge of the
molecular bar, and not the bar itself (see Fig.5 in Lundgren et al.~\cite{lundgren+04a}).
Furthermore, we should keep in mind that this may also affect the comparison with
the \ion{H}{i} data,
where 55\% of the total \ion{H}{i} emission within the Holmberg radius was detected
(Tilanus \& Allen \cite{tilanus+allen93}).

\section{Correlations between different tracers of the interstellar medium}

\subsection{The method}
\label{corr}

The correlations are determined using an algorithm that plots corresponding
values with linear or logarithmic axes (Nieten\ \cite{nieten01}), using
different gridding and weighting schemes. The results presented here
were obtained with a $1/\sigma$ weighting, where $\sigma$ is the rms
noise in each image. Only intensities with amplitudes above 5$\sigma$ were
used in order to avoid any influence of instrumental noise.
Only spatially independent points are selected. Selecting pixels separated
by 2.0 beam widths, at which point the Gaussian telescope beam overlaps by
only 2\%, is a good compromise between independence and maximal data
usage. Four
different hexagonal grids were used for each correlation to assess the
effect of gridding on the resulting correlations, and their mean value
was adopted. The difference in the correlation coefficient for slightly
offset hexagonal grids is typically a few percent.

We used the MIR, \ion{H}{i}, CO, H$\alpha$ and $\lambda$6~cm
radio continuum maps for this correlation study. The radio intensity
was separated into its thermal and
non-thermal components with help of a spectral index map between $\lambda$20~cm
(Sukumar \& Allen \cite{sukumar+allen89}) and $\lambda$6~cm, assuming constant
spectral indices for the thermal ($\alpha$=0.1) and non-thermal ($\alpha$=1.0)
components, as in Neininger et al. (\cite{neininger+93}).
 The optical images, broad-band R and narrow-band H$\alpha$, are from
Larsen \& Richtler (\cite{larsen+99}) and Lundgren et al.\
(\cite{lundgren+05}). All images were
smoothed to a common resolution of $23\arcsec$ ($\approx$500~pc).
The nuclear region was screened
with a diameter of $1\arcmin$, since we expect the nuclear area to
behave differently from the outer bulge, spiral arm and interarm regions.
We assume a relationship between the intensities $X$ and $Y$ at two wavelengths of the form:
\begin{equation}
Y  = 10^{\,\rm Const} \cdot X^{\rm Exp}
\end{equation}
The exponents, with the associated errors, as well as the
Pearson's correlation coefficients and Student-t values, are given in
Table~\ref{table:corr}.
Representative plots are presented in Fig.~\ref{fig:corr}.

\begin{figure*}
 \centerline{ \includegraphics[bb = 14 14 469 609,width=15cm]{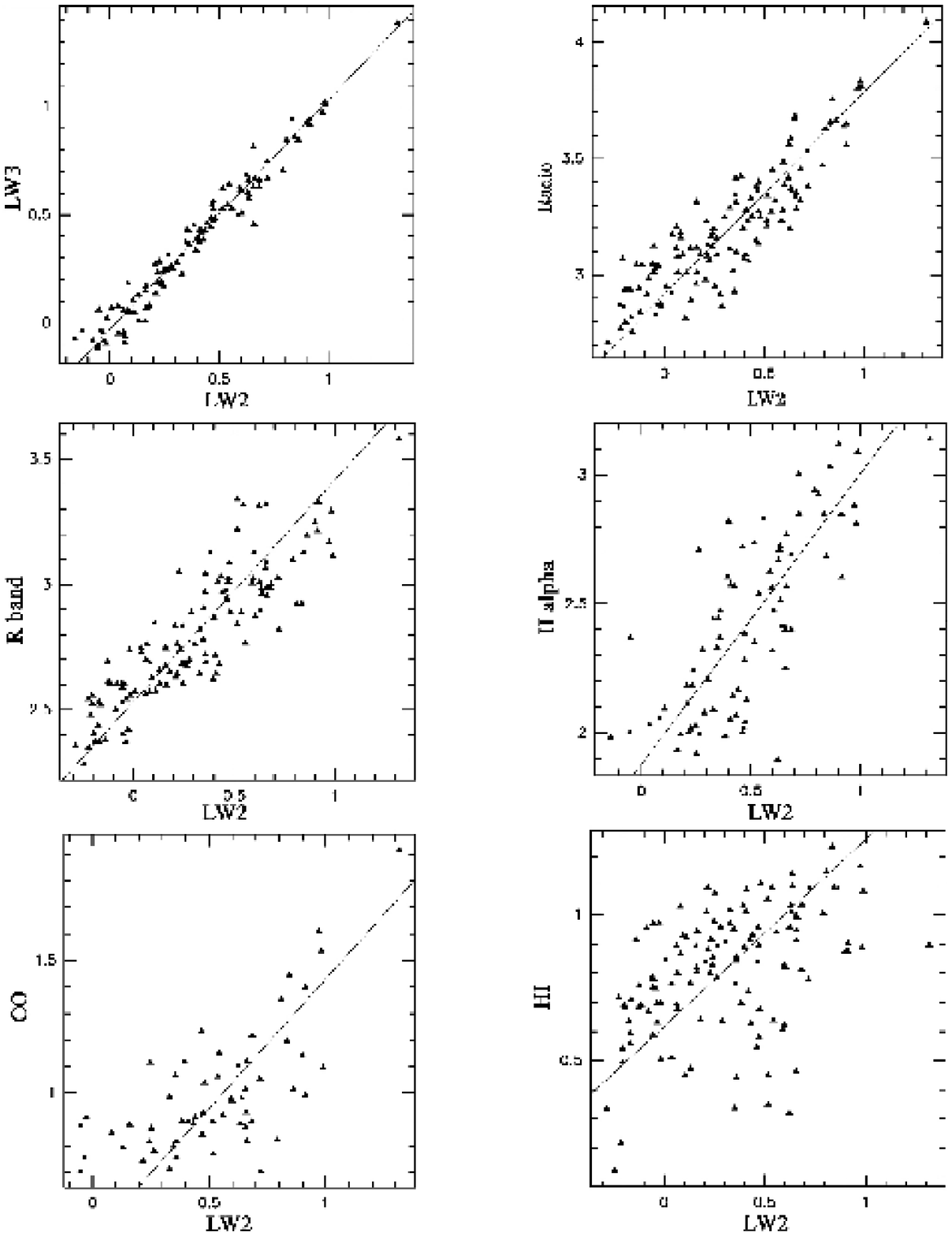}}
\caption{Correlations for different tracers of the interstellar medium in
M\,83 excluding the nuclear region ($0.5\arcmin \le R \le 5\arcmin$).
Images at a resolution of $23\arcsec$ were used. Points in the diagrams
represent independent pixels separated by two beam widths and having
signal-to-noise ratios of $\ge5$. We show the correlations between the
logarithms of intensities (in arbitrary units) in MIR filters
LW2 (centered at $\lambda$6.75\m) and LW3 ($\lambda$15\m), between LW2 and
$\lambda$6~cm radio continuum, LW2 and optical R band, LW2 and H$\alpha$,
LW2 and CO, and between LW2 and \ion{H}{i}.
}
\label{fig:corr}
\end{figure*}

\begin{table*}[tbh]
\caption[]{Correlations $X\longleftrightarrow Y$ between the logarithms of the intensities of various emissions in M\,83 at 23\arcsec\, resolution,
computed between 0.5 and 5\arcmin\, distance from the center in the galaxy plane.
The first value tabulated is the slope of the regression line (the exponent
in Eq.~(1)) and its error,
followed by the Pearson's correlation coefficient $r$, its error, and
the $t$ value of Student's t-test (see text for details).
}
\label{table:corr}
\[
\begin{tabular*}{15cm}[tbh]{lllllllll}
\hline
\hline
\noalign{\smallskip}
$X\,\,\,\,\,\setminus\,\,\,\,\,Y$  &15~$\mu$m\,\,\,\, &Radio\,\,\,\, &Thermal\,\, &Nonthermal &R band\,\, &H$\alpha$\,\,\,\,\, &\mbox{CO($J$=1--0)}\,\, &{H\,{\sc i}}\,\,\\
\noalign{\smallskip}
\hline
\hline
\noalign{\smallskip}
7~$\mu$m              & 1.05       & 0.86     & $\approx$0.8         & 0.8       & 0.89      & 1.26       & 0.9      & 0.65\\
                      & $\pm$0.03  & $\pm$0.12           & --        & $\pm$0.3  & $\pm$0.06 & $\pm$0.05 & $\pm$0.2  & $\pm$0.14\\
                      & 0.978 &  0.89         & 0.74      & 0.78      &0.89 & 0.84      & 0.70      & 0.43\\
                      & $\pm$0.004 & $\pm$0.02 & $\pm$0.07 & $\pm$0.04 & $\pm$0.02 & $\pm$0.03 & $\pm$0.07 & $\pm$0.07\\
                      & 48        & 21        & 7         & 13        & 21   & 15        & 7         & 5\\
\hline
\noalign{\smallskip}
15~$\mu$m             &     & 0.88      & $\approx$0.8 & 0.8       & 0.89      & 1.25      & 0.9       & 0.7\\
                      &       & $\pm$0.14 & --           & $\pm$0.3  & $\pm$0.09 & $\pm$0.06 & $\pm$0.2  & $\pm$0.2\\
                      &     & 0.88 & 0.73         & 0.82      & 0.87 & 0.80      & 0.72      & 0.26\\
                      & & $\pm$0.02  & $\pm$0.07    & $\pm$0.03 & $\pm$0.03 & $\pm$0.04 & $\pm$0.06 & $\pm$0.09\\
                      &     & 18   & 7            & 14        & 18   & 13         & 8         & 3\\
\hline
\noalign{\smallskip}
Radio                &            &     & 1.0       & 0.9       & 1.08      & 1.36      & 1.0       & 0.8\\
                     &            &       & $\pm$0.8  & $\pm$0.3  & $\pm$0.08 & $\pm$0.13 & $\pm$0.5  & $\pm$0.2\\
                     &            &
   & 0.79      & 0.90 & 0.85      & 0.68      & 0.77      & 0.53\\
                     &            &       & $\pm$0.06 & $\pm$0.02 & $\pm$0.02 & $\pm$0.06 & $\pm$0.05 & $\pm$0.06\\
                     &            &    & 8         & 21   & 19        & 9         & 9         & 7\\
\hline
\noalign{\smallskip}
Thermal               &            &           &         & 1.1       & 1.7       & 1.5       & 1.0       & 0.9\\
                      &            &           &           & $\pm$0.8  & $\pm$0.4  & $\pm$0.4  & $\pm$0.7  & $\pm$0.2\\
                      &            &           &         & 0.58      & 0.57      & 0.70      & 0.60      & 0.50\\
                      &            &           &           & $\pm$0.11 & $\pm$0.10 & $\pm$0.08 & $\pm$0.12 & $\pm$0.12\\
                      &            &           &         & 4         & 5         & 6         & 4         & 4\\
\hline
\noalign{\smallskip}
Nonthermal            &            &           &           &         & 1.13      & 1.5       & 1.1     &$\approx$0.9\\
                      &            &           &           &           & $\pm$0.14 & $\pm$0.2  & $\pm$0.6  & --\\
                      &            &           &           &        & 0.79      & 0.53      & 0.69      & $<$0.1\\
                      &            &           &           &           & $\pm$0.04 & $\pm$0.08 & $\pm$0.07 &  --\\
                      &            &           &           &         & 13        & 6         & 7         & $<$1\\
\hline
\noalign{\smallskip}
R band                &            &           &           &           &         & 1.3       & $\approx$1.0 & $\approx$0.5\\
                      &            &           &           &           &           & $\pm$0.2  & --      & -- \\
                      &            &           &           &           &         & 0.60      & 0.66      & 0.35\\
                      &            &           &           &           &           & $\pm$0.07 & $\pm$0.08 & $\pm$0.07\\
                      &            &           &           &           &         & 7         & 7         & 4\\
\hline
\noalign{\smallskip}
H$\alpha$             &            &           &           &           &           &         & 0.7       & 0.5\\
                      &            &           &           &           &           &           & $\pm$0.6  & $\pm$0.2\\
                      &            &           &           &           &           &         & 0.52      & 0.32\\
                      &            &           &           &           &           &           & $\pm$0.10 & $\pm$0.09\\
                      &            &           &           &           &           &         & 4         & 3\\
\hline
\noalign{\smallskip}
\mbox{CO($J$=1--0)}               &            &           &           &           &           &           &    & $\approx$-0.6\\
                      &            &           &           &           &           &           &           & --\\
                      &            &           &           &           &           &           &         &$<$0.1\\
                      &            &           &           &           &           &           &           & --\\
                      &            &           &           &           &           &           &         &$<$1\\
\hline
\hline
\end{tabular*}
\]
\end{table*}

The correlation between LW2 and LW3 is extremely good; the ratio
LW2/LW3 shows little variation within the disk (see
Sect.~\ref{lw2lw3corr} and the radial profiles in Fig~\ref{radial-profiles}).
The correlations of LW2 and LW3 with the
radio continuum and with the R band are likewise good, with a Pearson's correlation coefficient of
$\sim 0.88$, closely followed by that of H$\alpha$, having
correlation coefficients of 0.80 to 0.84.  The total radio continuum emission
is better correlated with the MIR emission than with each of the individual
thermal and non-thermal radio components. The CO emission does not
follow the MIR emission well (correlation coefficent of
0.7). The H$\alpha$ and the radio thermal emission are not well
correlated, probably an effect due to optical extinction. The
H$\alpha$ and the R band are also not well-correlated which is evident in
Fig~\ref{radial-profiles}.

\ion{H}{i} shows no significant correlation with any other tracer.
All these results agree with those found in NGC~6946 (Frick et al.\
\cite{frick+01}; Walsh et al.\ \cite{walsh+02}).

We also studied correlations between the images available
at $12\arcsec$ (260~pc) resolution which gave very similar results.
For example, the correlation coefficient of the LW2--LW3 correlation
was 0.95 and that of the LW3--radio continuum correlation was 0.84.

\subsection{The mid-infrared LW2 ($\lambda$6.75\m) -- LW3
($\lambda$15.0\m) correlation}
\label{lw2lw3corr}

Figures~\ref{lw21w3} and \ref{fig:corr} (top left panel) show an excellent
correlation between the two MIR broad-band filters. The correlation
coefficient is close to one ($0.978\pm0.004$) and the exponent
$1.05\pm0.03$, arguing for a nearly linear relationship.
In NGC~6946 (Walsh et al.\ \cite{walsh+02}), the correlation
coefficient and the exponent are almost identical to those in M~83.
The LW2/LW3 ratios (Table~\ref{cvf-spectra-tab}) vary
within M\,83 from 0.6 to 1.75. Since we screened the nuclear and inner
bulge region and required a signal-to-noise ratio $\gid 5$, thus
excluding a part of the interarm regions, we apparently pick up
ratios of 1.1 for the outer bulge and 0.8 for a region on the
spiral arm (with errors of $\sim 20\%-30\%$).

The overall LW2/LW3 ratio $\sim1$ is regarded to be typical
for spiral galaxies without active nuclei (Roussel et al.\
\cite{roussel+01a}, \cite{roussel+01b}; see Sect.~\ref{lw2lw3ref}).
Applying our findings to the case of spatially unresolved galaxies,
LW2/LW3 ratios $\sim 1$ for the total galaxy do not exclude the existence
of bursts of star formation in particular regions like the nucleus
or the inner spiral arms, where LW2/LW3$<1$
(see also Roussel et al.\ \cite{roussel+01c}).

\subsection{The mid-infrared -- radio correlation}
\label{mirradio}

The correlation between the integrated radio continuum and far-infrared
(FIR) fluxes of galaxies is one of the closest in astrophysics and has been
discussed over many years (e.g. Helou \& Bicay\ \cite{helou+bicay93};
Niklas \& Beck\ \cite{niklas+beck97}; Hoernes et al.\ \cite{hoernes+98}).
The mid-infrared -- radio correlation within M\,83 is also very tight
(Figs.~\ref{lw21w3}, \ref{lw2-over-radio}, \ref{fig:corr}).
The correlation coefficients between the LW2 and LW3 MIR bands and the radio intensity
are $0.89\pm0.02$ and $0.88\pm0.02$, respectively
(Table~\ref{table:corr}). The slopes of $0.86\pm0.12$ and $0.88\pm0.14$
are similar to that of the far-infrared -- radio correlation (Niklas \&
Beck\ \cite{niklas+beck97}; Hoernes et al.\ \cite{hoernes+98}) and
indicate that the same physical process is involved.
The warm dust consists of PAHs and grains of different sizes, emitting
in the mid- and far-infrared. Neither the individual thermal nor non-thermal radio
components are better correlated with MIR than the total radio emission.
The MIR emission, similar to the FIR emission, probably consists of
two components, a diffuse one (like that of the non-thermal radio emission)
and a clumpy one concentrated around star-forming regions (like that
of the thermal radio emission).

The explanation of the correlation may be threefold:

a) thermal emission,

b) cosmic-ray production in star-formation regions,

c) magnetic fields.

\medskip

a) If the radio continuum intensity in M\,83 is dominated by free-free
thermal emission, the correlation could be caused by the action of
star formation which heats the dust and ionizes the gas.  However,
thermal radio emission originates in the ionized gas phase, a phase
which primarily provides the dust continuum component of the LW3
emission and is also responsible for the excitation of the surrounding
PAHs dominating the LW2 emission. Furthermore, our separation of
thermal and non-thermal radio emission of M\,83, with the help of the
spectral index, reveals a dominance of non-thermal emission at all
radii. Thermal emission may dominate locally in star-forming regions,
but non-thermal processes are essential in order to understand the
correlation.

b) Star-forming regions are locations of type~II, Ib and Ic supernovae, and the
shock fronts of their remnants are probably cosmic-ray accelerators.
If the magnetic field strength were constant, the radio continuum image
would reflect the distribution of cosmic-ray electrons originating in
star-forming regions. The diffusion speed of cosmic rays is limited to
the Alfv\'en velocity ($\simeq 100$~km/s in the hot interstellar medium)
so that the particles propagate $\simeq 1\kpc$ within their lifetime of
$\simeq 10^7$~y. Under the given assumptions the radio continuum map
should look like a smoothed image of the star-forming regions (H$\alpha$),
which indeed is the case. Similarly, the MIR image traces the dust
in and around the star-forming regions.
A detailed analysis of this correlation ``scale
by scale'' with help of wavelets is underway (Beck et al.\ \cite{beck+05}).

c) However, a constant magnetic field strength in the disk of M\,83 is
physically unreasonable. Even small variations in field strength lead to
strong fluctuations in synchrotron intensity (see Sect.~\ref{radioinf}).
In this case any correlation with radio continuum emission needs to
consider magnetic fields.

The lack of large-scale correlations of radio continuum with neutral
\ion{H}{i} gas or with hot X-ray emitting gas from the disk (Ehle et al.\
\cite{ehle+98}) indicates that the warm neutral gas as well as the hot
ionized gas components are of little importance for the distribution
of magnetic fields.

We propose that {\it the interstellar
magnetic fields are connected to the gas clouds in which the dust is
embedded and they are probably anchored in their photoionized shells.}
This provides a relation between the number density $N$
of gas clouds and field strength $B$ of $B^2\propto N$.
A similar relation is obtained through equipartition between
the energy densities of turbulent cloud motion and magnetic fields,
as e.g. observed in NGC\,6946 (Beck\ \cite{beck04}).
The physical mechanism and time scale of this coupling has to be
investigated.

Our result also bears important consequences for the generation and
dynamics of magnetic fields. If the fields are anchored
in gas clouds, turbulent motions of gas clouds may cause
the ``$\alpha$--effect'' necessary for dynamo action
(Beck et al.\ \cite{beck+96}).

 Elbaz et al.\ (\cite{elbaz+02}) have demonstrated the existence of the global MIR-radio correlation for galaxies up to z=1, while Appleton et al.\  (\cite{appleton+04}) have found this correlation to remain out to z=2. Thus, globally for galaxies in the local universe and within individual galaxies, as we see here, as well as for cosmologically significant galaxies, this correlation remains a fundamentally important physical process.

\subsection{Comparison between mid-infrared and H$\alpha$}
\label{miralpha}

In our Galaxy, the PAH emission seems to be closely associated with
photodissociation shells and the surfaces of molecular clouds in the
vicinity of \ion{H}{ii} regions or exciting hot stars, while the VSG
continuum peaks inside \ion{H}{ii} regions (Siebenmorgen \&
Kr\"ugel\ \cite{sieben+krugel92}; Cesarsky et al.\ \cite{cesarsky+96b};
Verstraete et al.\ \cite{verstraete+96}). Both
regions are linked to star formation, which is also traced by
H$\alpha$ emission. From this, we would expect a correlation between
the MIR and H$\alpha$ emission (smoothed to the same resolution).
For M\,83 we find a correlation coefficient of $\simeq0.84$ and
$f_{\rm H\alpha}\sim f_{\rm MIR}^{1.3}$ (Table~5).
According to Fig.~\ref{lw2-over-halpha},
the MIR emission is more diffuse than
H$\alpha$ which is clumpy and concentrated toward the youngest
star-forming regions. This also explains the tight correlation
between the MIR and optical R band emissions.

While the radial profiles of H$\alpha$ and the LW2 and LW3 appear to
follow a similar pattern, there are significant relative spatial
differences near 2$\arcmin$, where the peak is more pronounced in
H$\alpha$ than in the MIR. Differences are also apparent in the outer
disk region, r$>$ 4' (Fig.~\ref{radial-profiles}). These effects
result in a slope of the regression exponent being 1.3
(Fig.~\ref{fig:corr}).
This exponent predicts, in
principle, the detection of decreasing MIR emission with respect to
the H$\alpha$ emission in the case of increasing radiation
fields. Given the large scatter of the individual values, this result
should be taken with care.

Roussel et al. (\cite{roussel+01b}) report on the relationship between
star-formation rates and mid-infrared emission in galactic disks for a
sample of 69 galaxies. Excluding the nuclear regions, they find, within
their errors, a linear relationship between the total MIR and H$\alpha$
emission in disks. The question arises whether, on a much larger scale
than the investigated one ($12\arcsec \cor 260\pc$), the MIR and H$\alpha$
emission correlate linearly. To check this, we made pixel
by pixel comparisons, requiring the same signal to noise ratio, with
artificially-reduced resolution. Reducing the resolution by a factor of
2, 4 or 8 still leads to a large scatter. With a resolution of
$96\arcsec \cor 2\kpc$ we find good correlations between {\it any}
tracers except \ion{H}{i} (correlation coefficients of $0.98\pm 0.01$,
but Student-t values of only $\simeq 10$). This reflects the obvious
fact that an extended disk exists in most spectral ranges (see
Fig.~\ref{lw21w3}). Thus correlation analysis of low-resolution data
can be misleading. A scale separation is required, as
demonstrated for NGC~6946 by Frick et al. (\cite{frick+01}).

\begin{figure}
\includegraphics[bb = -7 221 333 469,width=8.5cm,clip=]{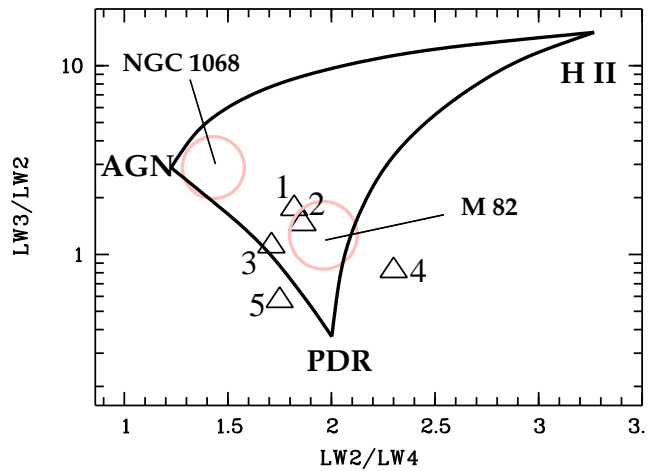}
\caption{\label{classification}
Classification diagram according to Laurent et al. (\cite{laurent+00}).
The positions of the extracted regions 1--5 of Sect.~\ref{regionsect}
are marked with triangles in this diagram.
}
\end{figure}

\subsection{Classification of M\,83 as a starburst galaxy with the help of
its MIR colours}

We calculated the LW3/LW2 and LW2/LW4 ratios for the five regions
presented in Sect.~\ref{regionsect} and Fig.~\ref{cvfima}. The ratios
are plotted in Fig.~\ref{classification}. The ratios can be used to build
a classification diagram (Laurent et al.\ \cite{laurent+00}), in which
the MIR emission from AGNs, photo-dissociation regions (PDR) and
\ion{H}{ii} regions are separated. The expectations for pure AGN, PDR
and \ion{H}{ii} regions and the results of Laurent et al.
(\cite{laurent+00}) for the AGN-dominated galaxy NGC~1068 and the
starburst dominated galaxy M~82 are indicated. Our measurements for the
nucleus, bulge and inner disk of M\,83 are in the vicinity of the M~82
measurements arguing for a high rate of star formation in these regions.
The regions 4 and 5 are slightly offset, which suggests a different
physical state. One should, however, keep in mind that the ratios for
the faint regions 4 and 5 are affected by larger errors ($\sim 30\%$).

\section{Summary}

M\,83 was observed with the ISOCAM in the 4{\m} to 18{\m}
band.
\begin{itemize}
\item
The LW2 and LW3 broad-band filter observations covered the entire optical
extent ($d_{25}=11\farcm5 \cor 15\kpc$ at $D=4.5$~Mpc assumed distance)
of the galaxy.
Both filters show bright emission from the nuclear and bulge regions,
star-formation regions at the tips of the bar and the inner spiral
arms. Emission from the outer spiral arms and interarm regions is
discernible with low intensity only. For LW2 ($\lambda$5{\m} to
8.50{\m}) and LW3 ($\lambda$12{\m} to 18{\m}), we measure an integral
flux  for M\,83 of 20~Jy ($1.6\times 10^{43}\ergs$) and 21~Jy
($6.6\times 10^{42}\ergs$), respectively. The central $3\arcmin\times
3\arcmin$ ($\cor 4\kpc \times 4\kpc$) region contributes $28\%$ and
$32\%$ to the total mid-infrared emission in the case of LW2 and LW3,
respectively.
\item
Strip maps, centered on the nucleus of M\,83, were taken in seven
narrow-band filters with scan directions from the north-east to the
south-west. The (14.0{\m}--16.0{\m})/(7.0{\m}--8.5{\m}) ratio is found
to be a good tracer of star-formation regions in the galaxy.
\item
The central $3\farcm2\times3\farcm2$ ($\cor 4\kpc\times 4\kpc$)
region of M\,83 was also observed in the spectral imaging mode. This
allowed us to extract $\lambda$4{\m} to 18{\m} spectra for 5 different
regions: the center, the inner and outer bulge, a mid-infrared emitting
on-arm region close to the south-western tip of the bar, as well as an
averaged interarm region spectrum. PAH bands are identified in all
regions at $\lambda$6.2{\m}, 7.7{\m}, 8.6{\m} and 11.3{\m}, as well as a
blended [Ne{\sc ii}] and PAH feature at $\simeq$12.8{\m}.
\item
Summed over the total ISOCAM MIR wavelength range of
$\lambda$4 to 18{\m}, the PAHs dominate in energy over the small grain
continuum emission.
While dust
continuum contributions make up 40\% of the total flux in the central region, they become almost negligible in the interarm
regions ($\la 10\%$). The contribution of atomic lines are small in
general, the brightest line being the [Ne{\sc ii}] line towards the
center. The ratio LW2/LW3 varies from $\simeq$0.6 (nucleus) to
$\simeq$1.8 (interarm regions).
\item
From our spectral imaging mode, we tested the PAH versus continuum
distribution for the broad-band filters LW2 and LW3. A great number of
galaxies have only been observed in these filter bands which were thought to
trace mainly the PAHs (LW2) and dust continuum (LW3). For LW2 we find
that PAH features indeed dominate the emission in the case of all
regions. For the LW3 band, the continuum emission dominates in the case
of the nucleus, bulge and spiral arms. For interarm or more diffuse
regions, however, PAH features are found to be the main source
of the LW3 emission.
\end{itemize}

New radio continuum (a combined $\lambda$6~cm VLA/Effelsberg map) and
H$\alpha$ data, which were used herein for a detailed quantitative
analysis, are presented. Both wavelengths show similar morphological
features as the mid-infrared emission does. Further tracers of the
interstellar medium, \mbox{CO($J$=1--0)}, \ion{H}{i}, diffuse X-rays,
as well as R band were compared to the mid-infrared images.

\begin{itemize}
\item
The radial profiles of the MIR, H$\alpha$, hard X-ray and
thermal radio continuum intensities in the range $2\arcmin$--$4\arcmin$ have
the same scale length of $1\farcm3\pm0\farcm1$ ($1.7\pm0.1$~kpc)
and steepen beyond 5~kpc radius.
The radial declines of the total radio continuum, CO and broad-band optical
emission are significantly flatter (scale lengths $1\farcm7$--$2\farcm2$).
\item
The morphology of the radio continuum and, to a large degree, H$\alpha$
resembles that of the mid-infrared. Correlation studies
between both two mid-infrared broad-bands and the radio continuum
intensities $I$ at $23\arcsec$ resolution show a nearly linear dependence
($I_\mathrm{radio} = \mathrm{const} \cdot\,I_\mathrm{LW2}^{0.86\pm0.12}$)
with a correlation coefficient of $0.89\pm0.02$. In the case of
H$\alpha$ we obtain $I_\mathrm{H\alpha} = \mathrm{const} \cdot
I_\mathrm{LW2}^{1.26\pm0.05}$ with a correlation coefficient of
$0.84\pm0.03$.
\item
The close relationship between mid-infrared and $\lambda$6~cm radio
continuum is discussed in the framework of different scenarios. A
possible implication of our findings is that the interstellar magnetic
fields are anchored in the photoionized shells of gas
clouds. This result would have important consequences for the
generation and dynamics of magnetic fields, e.g.,  the
``$\alpha$--effect'' necessary for dynamo action may be driven by
turbulent motions of gas clouds.
\end{itemize}

\begin{acknowledgements}
We wish to thank N. Neininger for help with the radio
observations, Ch. Nieten for providing and explaining his correlation
software, and T. Handa, G. Petitpas, R. Rand and R. Tilanus
for kindly providing their \mbox{CO($J$=1--0)}, CO($J$=4--3), CO($J$=1--0, S
arm) and \ion{H}{i} images.  We thank H. Roussel who provided the
preanalyzed ISOCAM CVF data cube. We thank E. Kr\"ugel for his insightful reading of the paper. We appreciate the careful reading of the manuscript by our referee and comments that helped improve the clarity of the presentation. Part of the MIR analysis was
carried out with non-standard CIA software written by P. Chanial,
H. Roussel and D. Reynaud. O. Laurent provided the source codes for the
spectral analysis of the CVF data. Our work made use of the archival
data base at the European Southern Observatory (ESO) and the NASA
Extragalactic Database (NED).
\end{acknowledgements}


\begin{thebibliography}{}

\bibitem[1987]{adamson+87}
   Adamson, A.~J., Adams, D.~J., \& Warwick, R.~S. 1987, MNRAS,
   224, 367

\bibitem[1989]{allamandola+89}
   Allamandola, L.~J., Tielens, A.~G.~G.~M., \& Barker, J.~R. 1989, ApJS, 71, 733

\bibitem[1999]{allamandola+99}
   Allamandola, L.~J., Hudgins, D.~M., \& Sandford, S.~A. 1999, ApJ, 511, L115

\bibitem[2004]{appleton+04}
   Appleton, P. N., Fadda, D., Marleau, F.R. et al. 2004, ApJ, 154, 147

\bibitem[1999]{arendt+99}
   Arendt, R.~G., Dwek, E., \& Mosley, S.~H. 1999, ApJ, 521, 234

\bibitem[1992]{athana92}
   Athanassoula, E. 1992, MNRAS, 259, 345

\bibitem[2002]{athey+02} Athey, A., Bregman, J., Bregman, J., Temi, P.,
   \& Sauvage, M. 2002, \apj, 571, 272

\bibitem[2002]{beck02}
   Beck, R. 2002, in: Disks of Galaxies: Kinematics, Dynamics and Pertubations,
   ed. E.~Athanassoula et al., ASP Conf. Ser. 275, 331

\bibitem[2004]{beck04}
   Beck, R. 2004, in: How Does the Galaxy Work?, eds. E.~J.~Alfaro et al.,
   (Dordrecht: Kluwer), 277

\bibitem[1996]{beck+hoernes96}
   Beck, R., \& Hoernes, P. 1996, Nat, 379, 47

\bibitem[1996]{beck+96}
   Beck, R., Brandenburg, A., Moss, D., Shukurov, A., \& Sokoloff, D.
   1996, ARAA, 34, 155

\bibitem[1999]{beck+99}
   Beck, R., Ehle, M., Shoutenkov, V., Shukurov, A., \& Sokoloff, D.
   1999, Nat, 397, 324

\bibitem[2005]{beck+05}
   Beck, R., Ehle, M., Frick, P., et al. 2005, in prep.

\bibitem[1988]{bland+tully88}
   Bland, J., \& Tully, B. 1988, Nat, 334, 43

\bibitem[2003]{bos+03} Boselli, A., Gavazzi, G., \& Sanvito, G. 2003, A\&A, 402, 37

\bibitem[1998]{boulanger+98}
   Boulanger, F., Boissel, P., Cesarsky, D., \& Ryter, C. 1998, A\&A, 339, 194

\bibitem[2002]{bresolin+kennicutt02}
   Bresolin, F., \& Kennicutt, R.~C. 2002, ApJ, 572, 838

\bibitem[1996]{cesarsky+96a}
   Cesarsky, C.~J., Abergel, A., Agnese, P., Altieri, B., \& Augueres,
   J.L. 1996a, A\&A, 315, 32

\bibitem[1996]{cesarsky+96b} Cesarsky, D., Lequeux, J., Abergel, A.,
   Perault, M., Palazzi, E., Madden, S., \& Tran, D. 1996b, \aap, 315, L309

\bibitem[2001]{chan+01}
   Chan, K.-W., Roellig, T.~L., Onaka, T., et al. 2001, ApJ, 546, 273

\bibitem[2000]{chanial+gastaud00}
   Chanial, P., \& Gastaud, R. 2000 (software available on email
   request via cir@discovery.saclay.cea.fr)

\bibitem[2000]{coulais+00} Coulais, A., \& Abergel, A. 2000, \aaps, 141, 533

\bibitem[2002]{crost+02}
   Crosthwaite, L.~P., Turner, J.~L., Buchholz, L., et al.
   2002, AJ, 123, 1892

\bibitem[2000]{dale+00}
    Dale, D.~A., Silbermann, N.~A., Helou, G., et al. 2000, AJ, 120, 583

\bibitem[1990]{desert+90}
   D\'esert, F.-X., Boulanger, F., \& Puget, J.~L. 1990, A\&A, 237, 215

\bibitem[1993]{deutsch+allen93}
   Deutsch, E.~W., \& Allen, R.~J. 1993, AJ, 106, 1812

\bibitem[1990]{dickey+lockman90}
   Dickey J.~M., \& Lockman F.~J., 1990, ARA\&A, 28, 215

\bibitem[2005]{dumke+05}
   Dumke, M., Thuma, G., Walsh, W., et al. 2005, A\&A, in prep.

\bibitem[1998]{ehle+98}
   Ehle, M., Pietsch, W., Beck, R., \& Klein, U. 1998, A\&A, 329, 39

\bibitem[2002]{elbaz+02}
   Elbaz, D., Cesarsky, C. J., Chanial, P., et al. 2002, A\&A, 384, 848

\bibitem[1998]{elmegreen+98}
   Elmegreen, D.~M., Chromey, F.~R., \& Warren, A.~R. 1998, AJ,
   116, 2834

\bibitem[2003]{forster+03}
   F\"orster-Schreiber, N.~M., Sauvage, M., Charmandaris, V., et al. 2003, A\&A, 399, 833

\bibitem[2001]{frick+01}
   Frick, P., Beck, R., Berkhuijsen, E.~M., \& Patrickeyev, I. 2001,
   MNRAS, 327, 1145

\bibitem[1991]{gallais+91}
   Gallais, P., Rouan, D., Lacombe, F., Tiph\'ene, D., \& Vauglin, I.
   1991, A\&A, 243, 309

\bibitem[2004]{galliano04}
   Galliano, F. 2004, PhD dissertation, University of Paris

\bibitem[1998]{genzel+98} Genzel, R., Lutz, D., Sturm, E., et al. 1998, \apj, 498, 579

\bibitem[2000]{genzel+cesar00} Genzel, R., \& Cesarsky, C.~J. 2000, \araa, 38, 761

\bibitem[1990]{handa+90}
   Handa, T., Nakai, N., Sofue, Y., Hayashi, M., \& Fujimoto, M. 1990, PASJ, 42, 1

\bibitem[1994]{handa+94} Handa, T., Ishizuki, S., \& Kawabe, R.\ 1994,
   in: Astronomy with Millimeter and Submillimeter
   Wave Interferometry, ASP Conf. Ser., 59, 341

\bibitem[1993]{helou+bicay93}
   Helou, G., \& Bicay, M.~D. 1993, ApJ, 415, 93

\bibitem[1996]{helou+96}
   Helou, G., Malhotra, S., Beichman, C., et al. 1996, A\&A, 315, L157

\bibitem[1998]{henning+98}
   Henning, T., Klein, R., Launhardt, R., Lemke, D. \& Pfau, W. 1998, A\&A, 332, 1035

\bibitem[1998]{hoernes+98}
   Hoernes, P., Berkhuijsen, E.~M., \& Xu, C. 1998, A\&A, 334, 57

\bibitem[1981]{hucht+bohnen81}
   Huchtmeier, W.~K., \& Bohnenstengel, H.-D. 1981, A\&A, 100, 72

\bibitem[2002]{hunt+02}
   Hunt, L.K., Giovanardi, C. \& Helou G. 2002, A\&A, 394, 873

\bibitem[1999]{immler+99}
   Immler, S., Vogler, A., Ehle, M., \& Pietsch, W. 1999, A\&A, 352, 415

\bibitem[1990]{jourdain+90}
   Jourdain de Muizon, M., d'Hendecourt, L.B. \& Geballe, T.~R. 1990, A\&A, 227, 526

\bibitem[1996]{kessler+96}
   Kessler, M., Steintz, J., Anderegg, M., et al. 1996, A\&A, 315, L27

\bibitem[2003]{krugel03}
   Kr\"ugel, E. 2003. The Physics of Interstellar Dust,
   (Bristol: Institute of Physics Publishing), 271

\bibitem[1996]{lagage+96}
   Lagage P.~O., Claret, A., Ballet, J., et al. 1996, A\&A, 315, L273

\bibitem[1999]{larsen+99}
   Larsen, S.~S., \& Richtler, T. 1999, \aap, 345, 59

\bibitem[2000]{laurent+00}
   Laurent, O., Mirabel, I.~F., Charmandaris, V., et al. 2000, A\&A, 359, 887

\bibitem[2001]{leech+01}
   Leech, K.~J., Metcalfe, L., \&  Altiere, B. 2001, MNRAS, 328, 1125

\bibitem[1999]{lehnert+99}
   Lehnert, M.~D., Heckman, T.~M., \& Weaver, K.~A. 1999, ApJ, 523, 575

\bibitem[1998]{lemke+98}
   Lemke, D., Mattila, K., \& Lehtinen, K. 1998, A\&A, 331, 742

\bibitem[1991]{lord+kenney91}
   Lord, S.~D., \& Kenney, J.~D.~P. 1991, ApJ, 381, 130

\bibitem[2004a]{lundgren+04a} Lundgren, A.~A., Wiklind, T., Olofsson, H.,
   \& Rydbeck, G. 2004a, \aap, 413, 505

\bibitem[2004b]{lundgren+04b} Lundgren, A.~A., Wiklind, T., Olofsson, H.,
   \& Rydbeck, G. 2004b, \aap, 422, 865

\bibitem[2005]{lundgren+05} Lundgren, A.~A., Olofsson, H. \& Wiklind, T., 2005, submitted to \aap,

\bibitem[2003]{lutz+03} Lutz, D., Sturm, E., Genzel, R., Spoon, H.~W.~W.,
   Moorwood, A.~F.~M., Netzer, H., \& Sternberg, A. 2003, \aap, 409, 867

\bibitem[2005]{madden+05}
   Madden, S.~C., Galliano, F., Jones. A. \& Sauvage, M. 2005, submitted to A\&A

\bibitem[1996]{mattila+96}
   Mattila, K., Lemke, D., Haikala, L.K., et al. 1996, A\&A, 315, L353

\bibitem[1991]{neininger+91}
   Neininger, N., Klein, U., Beck, R., \& Wielebinski, R. 1991, Nat, 352, 781

\bibitem[1993]{neininger+93}
   Neininger, N., Beck, R., Sukumar, S., \& Allen, R.~J. 1993, A\&A, 274, 687

\bibitem[2001]{nieten01}
   Nieten, C. 2001, PhD Thesis, University of Bonn

\bibitem[1997]{niklas+beck97}
   Niklas, S., \& Beck, R. 1997, A\&A, 320, 54

\bibitem[1990]{ohashi+90}
   Ohashi, T., Makishima, K., Tsuru, T., et al. 1990, ApJ, 365, 180

\bibitem[1997]{okada+97}
   Okada, K., Mitsuda, K., \& Dotani, T. 1997, PASJ, 49, 653

\bibitem[1998]{petit+wilson98}
   Petitpas, G.~R., \& Wilson, C.~D. 1998, ApJ, 503, 219

\bibitem[1984]{puget+84} Puget, J.~L., \& Leger, A. 1984, A\&A, 137, L5

\bibitem[1999]{rand+99}
   Rand, R.~J., Lord, S.~D., \& Higdon, J.~L. 1999, ApJ, 513, 720

\bibitem[2002]{rigopoulou+02} Rigopoulou, D., Kunze, D., Lutz, D., Genzel, R.,
   \& Moorwood, A.~F.~M. 2002, \aap, 389, 374

\bibitem[2001a]{roussel+01a}
   Roussel, H., Sauvage, M., Bosma A., et al. 2001a, A\&A, 369, 473

\bibitem[2001b]{roussel+01b}
   Roussel, H., Sauvage, M., Vigroux, L., et al. 2001b, A\&A, 372, 406

\bibitem[2001c]{roussel+01c}
   Roussel, H., Sauvage, M., Vigroux, L., et al. 2001c, A\&A, 372, 427

\bibitem[1995]{ryder+95}
   Ryder, S.~D., Hungerford, A., Dopita, M.~A., et al. 1995,
   in: The Opacity of Spiral Disks, ed. J.~I. Davies \& D. Burstein
   (Dordrecht: Kluwer), 359

\bibitem[1994]{ryder+94}
   Ryder, S.D., Dopita, M. A. 1994, \apj, 430, 142

\bibitem[2004]{sakamoto+04}
    Sakamoto, K., Matsushita, S., Peck, A.B., Wiedner, M. C., Iono, D. 2004, ApJ, 616, L59

\bibitem[1998]{schlegel+98}
   Schlegel, D. J., Finkbeiner, D. P., \& Davis, M. 1998, ApJ, 500, 525

\bibitem[1992]{schulz+wegner92}
   Schulz, H., \& Wegner, G. 1992, A\&A, 266, 167

\bibitem[1993]{schutte+93}
   Schutte, W.~A., Tielens, A.~G.~G.~M., \& Allamandola, L.~J. 1993, ApJ, 415, 397

\bibitem[1990]{sellgren+90}
   Sellgren, K., Luan, L., \& Werner, M.~W. 1990, ApJ, 359, 384

\bibitem[1992]{sieben+krugel92}
   Siebenmorgen, R., \& Kr\"ugel, E. 1992, A\&A, 259, 614

\bibitem[1994]{sofue+94}
   Sofue Y., \& Wakamatsu T. 1994, AJ, 107, 1018

\bibitem[2002]{soria+wu02}
   Soria, R., \& Wu, K. 2002, A\&A, 384, 99

\bibitem[2003]{soria+wu03}
   Soria, R., \& Wu, K. 2003, A\&A, 410, 53

\bibitem[2000]{sturm+00} Sturm, E., Lutz, D., Tran, D.,  et al. 2000, \aap, 358, 481

\bibitem[2002]{sturm+02} Sturm, E., Lutz, D., Verma, A., et al. 2002, \aap, 393, 821

\bibitem[1989]{sukumar+allen89} Sukumar, S., \& Allen, R.~J. 1989, Nat, 340, 537

\bibitem[2000]{thatte+00}
   Thatte, N., Tecza, M., \& Genzel, R. 2000, A\&A, 364, L47

\bibitem[2003]{thim+03}
   Thim, F., Tammann, G.~A., Saha, A., et al. 2003, ApJ, 590, 256

\bibitem[1993]{tilanus+allen93}
   Tilanus, R.~P.~J., \& Allen, R.~J. 1993, A\&A, 274, 707

\bibitem[1985]{trinchieri+85}
   Trinchieri, G., Fabbiano, G., \& Palumbo, G.~G.~C. 1985, ApJ, 290, 96

\bibitem[2000]{uchida+00}
   Uchida, K.~I., Sellgren, K., Werner, M.~W., \& Houdashelt, M.~L. 2000, ApJ, 530, 817

\bibitem[2003]{verma+03}
   Verma, A., Lutz, D., Sturm, E., et al. 2003, A\&A, 403, 829

\bibitem[2002]{vermeij+02}
   Vermeij, R., Peeters, E., Tielens, A.~G.~G.~M., \& van der Hulst, J.M. 2002, A\&A, 382, 1042

\bibitem[1996]{verstraete+96}
   Verstraete, L., Puget, J.~L., Falgarone, E., et al. 1996, A\&A, 315, L337

\bibitem[2001]{verstraete+01}
   Verstraete, L., Pech, C., Moutou, C., et al. 2001, A\&A, 372, 981

\bibitem[1999]{vigroux+99}
   Vigroux, L., Charmandaris, V., Gallais, P., et al. 1999, in: The
   Universe as Seen by ISO, ed. P. Cox \& M.~F. Kessler, ESA-SP 427, 805

\bibitem[2002]{walsh+02}
   Walsh, W., Beck, R., Thuma, G., et al. 2002, A\&A, 388, 7

\bibitem[2004]{xilouris+04}
   Xilouris, E.~M., Madden, S.~C., Galliano, F., Vigroux, L.,
   \& Sauvage, M. 2004, \aap, 416, 41


\end{thebibliography}
\end{document}